\documentclass[twocolumn,english,letterpaper,superscriptaddress,
prb,aps,showpacs]{revtex4}

\usepackage[T1]{fontenc}
\usepackage[latin1]{inputenc}
\usepackage{amsmath}
\usepackage{graphicx}
\usepackage{amssymb}

\usepackage{color} 
\usepackage{graphicx}
\usepackage{bm}
\usepackage{amsmath}
\makeatletter
\begin{document}
\newcommand{\kv}{{\bm k}}
\newcommand{\qv}{{\bm q}}
\newcommand{\nv}{{\bm n}}
\newcommand{\pv}{{\bm p}}
\newcommand{\rv}{{\bm r}}
\newcommand{\vare}{\varepsilon}
\newcommand{\beqa}{\begin{eqnarray}}
\newcommand{\eeqa}{\end{eqnarray}}
\newcommand{\beq}{\begin{equation}}
\newcommand{\eeq}{\end{equation}}
\newcommand{\bars}{\bar{\sigma}}
\newcommand{\upa}{\uparrow}
\newcommand{\doa}{\downarrow}
\newcommand{\Gt}{\tilde\Gamma_2}
\newcommand{\Gi}{\tilde\Gamma_1}

\title{Spin and charge transport induced by gauge fields in a ferromagnet}
\author{Junya Shibata}
\email{j_shibata@toyo.jp}
\affiliation{Faculty of Science and Engineering, 
Toyo University, Kawagoe, Saitama, 350-8585, Japan}
\author{Hiroshi Kohno}
\email{kohno@mp.es.osaka-u.ac.jp}
\affiliation{Graduate School of Engineering Science, Osaka University,
Toyonaka, Osaka 560-8531, Japan}%
\date{\today}

\begin{abstract}
We present a microscopic theory of spin-dependent motive force (``spin motive force'') 
induced by magnetization dynamics 
in a conducting ferromagnet, by taking account of spin relaxation of conduction electrons. 
 The theory is developed by calculating spin and charge transport driven by two kinds of gauge fields; 
one is the ordinary electromagnetic field $A^{\rm em}_{\mu}$, 
and the other is the effective gauge field $A^{z}_{\mu}$ induced by dynamical magnetic texture. 
 The latter acts in the spin channel and gives rise to a 
spin motive force. 
 It is found that the current induced as a linear response to $A^{z}_{\mu}$ 
is not gauge-invariant in the presence of spin-flip processes. 
 This fact is intimately related to the non-conservation of spin via Onsager reciprocity, 
so is robust, but indicates a theoretical inconsistency. 
 This problem is resolved by considering the time dependence of spin-relaxation source terms   
in the \lq\lq rotated frame'', 
as in the previous study on Gilbert damping 
[J.~Phys.~Soc.~Jpn.~{\bf 76}, 063710 (2007)]. 
 This effect restores the gauge invariance while keeping spin non-conservation. 
 It also gives a dissipative spin motive force 
expected as a reciprocal to the dissipative spin torque (\lq\lq $\beta$-term''). 
\end{abstract}  

\maketitle

\section{Introduction}
 Manipulation of magnetization by electric currents \cite{{Berger78},{Slonczewski},{Berger96}} 
has been studied intensively 
for a decade because of promising spintronic applications.\cite{Parkin07} 
 Among them, it was demonstrated theoretically\cite{TKS08} and experimentally
\cite{Ono09} 
that an electric current in a conducting ferromagnet can 
drive magnetic textures such as domain walls and vortices. 
 This is understood as due to spin torques that a current exerts on magnetization 
through a microscopic exchange interaction.
 They include the spin-transfer torque, 
\cite{BJZ98,Ansermet04,Li04,Nakatani04} 
which is based on the conservation of total angular momentum,  
and its dissipative correction called $\beta$-term,
\cite{{Zhang04},{TNMS05},{TSBB06},{KTS06},{KS07},{Duine07_PRB},{PT07}}  
which arises in the presence of 
spin-relaxation processes in the electron system.

In 1986, Berger predicted a reciprocal effect 
that a moving domain wall accompanied by a periodic rotation of magnetization generates an electromotive force,   
in analogy with the Josephson effect of superconductivity. \cite{Berger86} 
 This effect is now understood as a motive force acting in spin channel, hence called spin motive force, 
\cite{Volovik87,Stern92,BM07,Duine08,Tserkovnyak08,Saslow07,Ohe09,YBKXNTE09,Yang10}  
which drives majority-spin and minority-spin electrons in mutually opposite directions. 
 It is also understood to arise from a time-dependent magnetic texture in general.  
 Recently, it was experimentally detected  by Yang {\it el al.}\cite{YBKXNTE09} 
for a vortex wall in a ferromagnetic nanowire. 
 Similar phenomena have also been studied in systems with interface or nanoparticles. \cite{BTBH02,Wang06,Costache06,THT10,SUMT06,Tanaka09}

 A theoretical framework for studying spin motive force in ferromagnets 
was presented by Volovik, \cite{Volovik87} or earlier by Korenmann {\it et al.} \cite{KMP77} 
 To treat electrons in a spin (or magnetization) texture, they introduced a local spin frame 
whose quantization axis coincides with the local spin direction,\cite{note1} ${\bm n}$;
then there arises naturally an effective U(1) gauge field, $A^{z}_{\mu}$, 
acting in electron's spin channel, 
which gives rise to an effective \lq electric' field 
\cite{{Volovik87},{Tserkovnyak08}}
\beqa 
\label{effective-field}
 E_{{\rm s},i}^0 
&=& \frac{\hbar}{e} (\partial_{i}A^{z}_{0}-\partial_{0}A^{z}_{i})
 = \frac{\hbar}{2e} \nv \!\cdot\! (\partial_{i}\nv \times \dot{\nv}), 
\eeqa
or a spin motive force, ${\bm F}_{\rm s} = -e{\bm E}_{\rm s}$    
($-e$\,: electron charge). 
 Recently, it was pointed out that it acquires a dissipative correction\cite{Duine08,Tserkovnyak08}
\beqa
\label{diss-effective-field}
 E_{{\rm s},i}^{\rm dis} = \beta \frac{\hbar}{2e} \, \dot{\nv} \!\cdot \partial_{i} \nv , 
\eeqa
in the presence of spin relaxation of conduction electrons. 
 The total field is then given by 
${\bm E}_{\rm s} = {\bm E}_{\rm s}^0 + {\bm E}_{\rm s}^{\rm dis}$. 
 These two terms are reciprocals to the spin-transfer torque and the spin torque $\beta$-term, respectively, 
\cite{{Saslow07},{Duine08},{Tserkovnyak08}} 
and the dimensionless parameter $\beta$ is the same as that of spin torque.  
\cite{{Zhang04},{TNMS05},{TSBB06},{KTS06},{KS07},{Duine07_PRB},{PT07}}

 A spin motive field ${\bm E}_{\rm s}$ induces an electric current 
\beqa
\label{intro-charge-current}
 {\bm j} = \sigma_\uparrow {\bm E}_{\rm s} +  \sigma_\downarrow  (-{\bm E}_{\rm s}) 
= \sigma_{\rm s} {\bm E}_{\rm s} , 
\eeqa
where $\sigma_\upa$ ($\sigma_\doa$) is a conductivity of majority- (minority-) spin electrons, 
and $\sigma_{\rm s} = \sigma_{\upa}-\sigma_{\doa}$ is the \lq spin conductivity'. 
 In most theoretical studies, this relation is used to identify a spin motive force.
\cite{{Saslow07},{Duine08},{Tserkovnyak08}}  
 In the presence of spin-orbit coupling, it induces in addition a charge Hall current, 
$\sigma_{\rm SH} {\bm n} \times {\bm E}_{\rm s}^0$, where $\sigma_{\rm SH}$ is a spin Hall conductivity, \cite{SK09}  
and as a reciprocal to this, a spin Hall current induced by external electric field will exert 
a spin-transfer torque. \cite{SK10} 
 Enhancement of magnetization damping due to induced spin current was also discussed.\cite{{Zhang09},{Zhang10}}

 The purpose of this paper is to develop a microscopic theory of spin motive force 
basing on the gauge field mentioned above.  
 For this, we found it instructive to treat spin and charge channels 
in parallel. 
 We thus study spin and charge transport induced by two kinds of gauge fields, 
one acting in charge channel (ordinary electromagnetic field) 
and the other acting in spin channel (spin motive field). 
  Particular attention is paid to the effects of spin relaxation of conduction electrons.

In the first part of this paper, 
we study spin and charge transport in a uniformly magnetized state induced by an ordinary electromagnetic field. 
 Our calculation is equivalent to the well-studied two-current model,
 {\cite{Mott64, Fert-Campbell68,Son87,Valet-Fert93}}
but some interesting crossover is pointed out in diffusion modes.

In the second part, 
we study a spin motive force by calculating electric and spin currents induced by magnetization dynamics. 
 We encounter a difficulty that 
the current induced as a linear response to the effective gauge field $A^{z}_{\mu}$ 
contains gauge non-invariant terms in the presence of spin-flip processes. 
 This difficulty is resolved by noting that there is another 
contribution from the source term of spin relaxation,    
as realized in the study of Gilbert damping. \cite{KS07}  
 We also found that such additional contribution reproduces the dissipative spin motive force. 

 Such additional contributions may look tricky, but their necessity can be understood on general grounds. 
 In the present gauge-field formalism, in which spin and charge channels are treated equally, 
spin conservation and gauge invariance (in the spin channel) are 
equivalent at the linear-response level because of Onsager reciprocity. 
 However, the former is violated by spin-flip processes whereas the latter should always hold   
in order for the theory to be consistent. 
 These contradictory aspects can only be reconciled by some additional contributions.

The paper is organized as follows. 
 After describing a model in Sec.~II,  
we examine in Sec.~III the density and current response to the ordinary electromagnetic field, $A^{\rm em}_{\mu}$.
 Here the magnetization is assumed to be static and uniform. 
 In Sec. IV, we consider the case that the magnetization varies in space and time. 
 By introducing another gauge field, $A^{z}_{\mu}$, which expresses the effects of magnetic texture and dynamics, 
we examine the density and current within the linear response to $A^{z}_{\mu}$, with an unpleasant, 
gauge-dependent result.  
 This problem is resolved in Sec.~V, where a dissipative correction to spin motive force is also obtained.
 Results and discussion are given in Sec.~VI, and summary is given in Sec.~VII.  
 Calculational details are given in Appendices.

\section{Model}
 We consider a ferromagnetic conductor 
consisting of conducting $s$-electrons and localized $d$-spins. 
We assume that 
the $s$-electrons are degenerate free electrons subject to impurity scattering, 
and localized $d$-spins are classical, 
which are mutually coupled via the $s$-$d$ exchange interaction. 
The Lagrangian for $s$-electrons is given by 
$L = L_{\rm el} -H_{\rm sd}$: 
\beqa
&&L_{\rm el} = \int d\rv~ c^{\dagger}\left[i\hbar
\frac{\partial}{\partial t}+
\frac{\hbar^2}{2m}\nabla^{2}
+\vare_{\rm F}-V_{\rm imp}\right]c, 
\label{Lagrangian-el1}\\
&&H_{\rm sd} = -M\int d\rv~ {\nv}
\cdot(c^{\dagger}{\bm \sigma}c)_{x}, 
\label{Hamiltonian-sd1}
\eeqa
where $c^{\dagger}(x)=(c^{\dagger}_{\uparrow}(x),c^{\dagger}_{\downarrow}(x))$ 
is the electron creation operator 
at $x=(t,\rv)$, $\vare_{\rm F}$ is the Fermi energy, 
$M$ is the $s$-$d$ exchange coupling constant, 
${\bm n}$ is a unit vector representing the direction of $d$-spin,\cite{note1} 
and ${\bm \sigma}$ is a vector of Pauli spin matrices. 
 The impurity potential is modeled by 
\beqa
V_{\rm imp}(\rv) = u_{\rm i}\sum_{i}\delta(\rv-{\bm R}_{i})
+u_{\rm s}\sum_{j}\delta(\rv-{\bm R}'_{j}){\bm S}_{j}\cdot{\bm \sigma}, 
\eeqa
where $u_{\rm i}$ and ${\bm R}_{i}$ are the strength and position 
of normal impurities, which introduce momentum relaxation processes, 
and $u_{\rm s}$ and ${\bm R}'_{j}$ are those of quenched magnetic impurities 
with spin ${\bm S}_{j}$, which introduce spin-relaxation processes. 
\cite{{KTS06},{KS07}} 
 We take a quenched average for the impurity spin 
direction as $\overline{S^{\alpha}_{i}}=0$ and \cite{subscripts}
\begin{equation}
  \overline{S_i^\alpha S_j^\beta} 
= \delta_{ij} \delta^{\alpha\beta} \times \left\{ \begin{array}{cc} 
  \overline{S_\perp^2} & (\alpha, \beta = x,y) \\ 
  \overline{S_z^2}     & (\alpha, \beta = z) 
  \end{array} \right.
\end{equation}
as well as for the impurity positions, ${\bm R}'_{i}$ and ${\bm R}'_{j}$. 
 When the magnetization is uniform and static, ${\nv} = \hat{z}$, 
the impurity-averaged Green's function is given by 
\beqa
G_{\kv\sigma}(z) = \frac{1}{z-\vare_{\kv}+\vare_{{\rm F}\sigma}+
i\gamma_{\sigma}{\rm sgn}({\rm Im}z)}, 
\eeqa
where  
$\kv$ is a wavevector, 
$\vare_{\kv} = \hbar^{2}\kv^{2}/2m$, and 
$\vare_{{\rm F}\sigma}=\vare_{\rm F}+\sigma M$. 
 The subscript $\sigma = \uparrow,\downarrow$ 
represents the majority and minority spins, respectively,  
and corresponds to $\sigma = +1,-1$ 
in the formula (and to $\bar{\sigma}= \downarrow, \uparrow$ or $-1$, $+1$). 
 Treating $V_{\rm imp}$ as perturbation, the damping rate $\gamma_{\sigma}$ is evaluated 
in the first Born approximation as 
\beqa
\gamma_{\sigma} = \frac{\hbar}{2\tau_{\sigma}} = 
\pi ( \Gi \nu_{\sigma} + \Gt \nu_{\bars}), 
\label{damping}
\eeqa 
where $\nu_{\sigma}=mk_{{\rm F}\sigma}/2\pi^{2}\hbar^{2}$ 
is the density of states at $\vare_{{\rm F}\sigma}$
with $k_{{\rm F}\sigma} = \sqrt{2m\vare_{{\rm F}\sigma}}/\hbar$ and 
\beqa
&&\Gi=n_{\rm i}u^{2}+n_{\rm s}u^{2}_{\rm s} \overline{S_z^2}, 
\label{Gamma1}\\
&&\Gt=2 n_{\rm s}u^{2}_{\rm s} \overline{S_\perp^2} , 
\label{Gamma2}
\eeqa 
with $n_{\rm i}$ and $n_{\rm s}$ 
being the concentration of normal and magnetic impurities, 
respectively. 
 The first and second terms in Eq.~(\ref{damping}) 
come from spin-conserving and spin-flip scattering processes, respectively.

 In this paper, we assume $\gamma_{\sigma} \ll \vare_{{\rm F}\sigma}$ 
and focus on diffusive transport 
induced by slowly-varying external perturbations (electromagnetic fields or time-dependent magnetic texture). 
 Let $q$ and $\omega$ be wavenumber and frequency of the perturbation, and define 
\beqa
 X_{\sigma} = (D_{\sigma} q^{2}-i\omega) \tau_{\sigma} , 
\label{eq:X}
\eeqa
with a diffusion constant $D_{\sigma}$. 
 Then  
our assumption throughout the paper is expressed as $\gamma_{\sigma} \ll \vare_{{\rm F}\sigma}$ and $|X_{\sigma}| \ll 1$.

\section{Spin and Charge transport in uniformly magnetized state}
\subsection{Linear response to electromagnetic field}
Let us examine the density and current response 
in the charge channel, $j_\mu = (\rho, {\bm j})$, and spin channel, 
$j_{{\rm s},\mu} = (\rho_{\rm s}, {\bm j}_{\rm s})$, 
to the external electromagnetic field, 
$A^{\rm em}_{\mu} = (-\phi^{\rm em}, {\bm A}^{\rm em})$.\cite{subscripts,note2} 
 Here $\phi^{\rm em}$ and ${\bm A}^{\rm em}$ are scalar and vector
potentials, respectively, and the time and space components of the four currents are given by 
\beqa
&{}& \hskip -5mm 
 \rho = -ec^{\dagger}c \ (\ = j^{(0)}_{0}),
\\
&{}& \hskip -5mm 
  {\bm j} = {\bm j}^{(0)} + \frac{e}{m} \rho {\bm A}^{\rm em}, 
\ \ \ 
{\bm j}^{(0)} = \frac{-e\hbar}{2mi} \, 
 c^{\dagger} \overset{\leftrightarrow}{\nabla} c, 
\\
&{}& \hskip -5mm 
  \rho_{\rm s} = -ec^{\dagger}\sigma^{z}c \ (\ =j^{(0)}_{{\rm s},0}) ,
\\
&{}& \hskip -5mm 
  {\bm j}_{\rm s} = {\bm j}_{\rm s}^{(0)} + \frac{e}{m} \rho_{\rm s} {\bm A}^{\rm em}, 
\ \ \ 
{\bm j}^{(0)}_{\rm s} = \frac{-e\hbar}{2mi} \, 
  c^{\dagger} \sigma^{z} \overset{\leftrightarrow}{\nabla} c , 
\eeqa 
with $c^{\dagger}\overset{\leftrightarrow}{\nabla}c = c^{\dagger} \nabla c - (\nabla c^\dagger )c$. 
 We have defined $\rho_{\rm s}$ and ${\bm j}_{\rm s}$ to have the same dimensions as 
$\rho$ and ${\bm j}$, respectively. 
 The coupling to the external fields is given by 
\beqa
H_{\rm em} &=& \int d\rv~
( \, \rho \, \phi^{\rm em}-{\bm j}^{(0)} \!\cdot\! {\bm A}^{\rm em})
\nonumber\\
&=& -\int d\rv~j_{\mu}^{(0)}A_{\mu}^{\rm em} .
\label{H-em}
\eeqa
 The currents, $j_{\mu}$ and $j_{{\rm s},\mu}$, are evaluated in the linear response to $A^{\rm em}_{\mu}$ as 
\beqa
\langle j_{\mu}(\qv) \rangle_{\omega} 
&=& e^{2}K_{\mu\nu}^{\rm cc}(\qv,\omega +i0)A^{\rm em}_{\qv,\nu}(\omega),
\label{current-1}\\
\langle j_{{\rm s},\mu}(\qv) \rangle_{\omega}
&=& e^{2}K^{\rm sc}_{\mu\nu}(\qv,\omega +i0)A^{\rm em}_{\qv,\nu}(\omega), 
\label{spin-current-1}
\eeqa
where $A^{\rm em}_{\qv,\nu}(\omega)$ 
is a Fourier component of $A^{\rm em}_\nu (x)$ . 
 The response functions $K_{\mu\nu}^{\rm cc}$ and $K^{\rm sc}_{\mu\nu}$ are obtained from   
\beqa
 e^{2}K^{\rm cc}_{\mu\nu}(\qv,i\omega_{\lambda}) &=&
\int_{0}^{1/T} d\tau~e^{i\omega_{\lambda}\tau}
 \langle {\rm T}_{\tau} j_{\mu}^{(0)}(\qv,\tau)j_{\nu}^{(0)}(-\qv) \rangle
\nonumber\\
&+& \frac{e}{m}\langle \rho \rangle \, \delta_{\mu\nu} (1-\delta_{\nu0}), 
\label{cc-1}
\\
 e^{2}K_{\mu\nu}^{\rm sc}(\qv,i\omega_{\lambda}) 
&=& \int_{0}^{1/T} d\tau~e^{i\omega_{\lambda}\tau}
    \langle {\rm T}_{\tau}
    j_{{\rm s},\mu}^{(0)}(\qv,\tau)j_{\nu}^{(0)}(-\qv) \rangle
\nonumber\\
&+& \frac{e}{m}\langle\rho_{\rm s}\rangle \, 
    \delta_{\mu\nu}(1-\delta_{\nu0}), 
\label{sc-1}
\eeqa
by the analytic continuation, $i\omega_\lambda \to \hbar \omega + i0$, 
where $\omega_{\lambda} = 2\pi \lambda T$ ($\lambda\,$: integer) 
is a bosonic Matsubara frequency.   
 In this paper, we focus on absolute zero, $T=0$. 
 The average $\langle\cdots\rangle$ is taken in the equilibrium state 
determined by $L$. 
 The Fourier components of the currents are given by 
\beqa
j_{\mu}^{(0)}(\qv) &=& -e\sum_{\kv,\sigma}v_{\mu}
c^{\dagger}_{\kv_{-},\sigma}c_{\kv_{+},\sigma},\\
j_{{\rm s},\mu}^{(0)}(\qv) &=& -e\sum_{\kv,\sigma}
\sigma v_{\mu}c^{\dagger}_{\kv_{-},\sigma}c_{\kv_{+},\sigma}
\eeqa
with 
\beqa
v_{\mu} = \left\{ \begin{array}{ll}
                   1 & (\mu =0)\quad  \\
                   \hbar k_{i}/m & (\mu = i = 1,2,3) \quad
                  \end{array} \right. 
\eeqa
and $\kv_{\pm} = \kv \pm \qv/2$.

\begin{figure}
\includegraphics[scale=0.26]{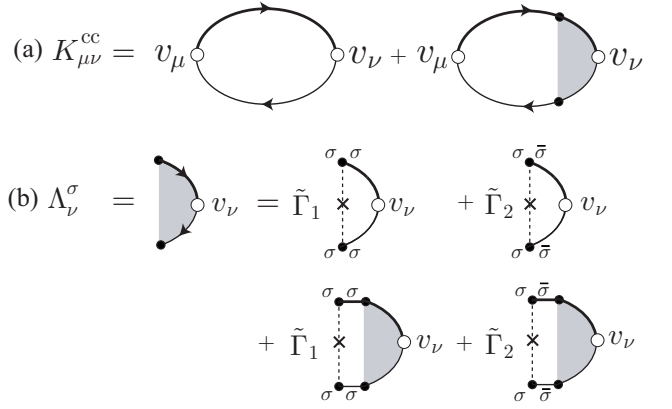}
\caption{(a) Diagrammatic expression of $K^{\rm cc}_{\mu\nu}$. 
The thick (thin) solid line represents an electron line 
carrying Matsubara frequency $i\vare_{n}+ i\omega_{\lambda}$ $(i\vare_{n})$. 
The shaded part represents the vertex function, $\Lambda^{\sigma}_{\nu}$. 
(b) Dyson equation for $\Lambda^{\sigma}_{\nu}$. 
 The dotted lines represent impurity scattering, 
either with ($\Gt$) or without ($\Gi$) spin-flip scattering.} 
\label{vertex_1}
\end{figure}

 The response functions are evaluated with the ladder-type vertex corrections \cite{AGD} [Fig.~\ref{vertex_1}(a)].  
Deferring the details to Appendix A, we give the results in the next subsection. 
 The results are concisely expressed with the quantities 
\beqa
Y_{\sigma} &=& D_{\sigma} q^{2}-i\omega ,
\\ 
Z &=& Y_{\upa}Y_{\doa} + 2\pi \Gt\langle Y \nu \rangle ,
\eeqa
and a notation, $\langle \cdots \rangle$, meaning to sum over $\sigma = \uparrow, \downarrow$; 
for example,  
$\langle\nu\rangle = \nu_{\upa} + \nu_{\doa}$, 
$\langle \sigma \nu\rangle = \nu_{\upa} - \nu_{\doa}$, 
$\langle D\nu \rangle = D_{\upa}\nu_{\upa} + D_{\doa}\nu_{\doa}$, and 
$\langle \sigma D\nu \rangle = D_{\upa}\nu_{\upa} - D_{\doa}\nu_{\doa}$. 
 By defining $(\bar Y)_\sigma = Y_{\bar\sigma}$, 
we may also use 
$\langle D\nu \bar{Y}\rangle 
= D_{\uparrow}\nu_{\uparrow}Y_{\downarrow} + D_{\downarrow}\nu_{\downarrow}Y_{\uparrow}$ 
and 
$\langle \sigma D\nu \bar{Y}\rangle 
= D_{\uparrow}\nu_{\uparrow}Y_{\downarrow} - D_{\downarrow}\nu_{\downarrow}Y_{\uparrow}$.

\subsection{Result}
\subsubsection{Charge channel}
 The response functions $K^{\rm cc}_{\mu\nu}(\qv,\omega+i0)$
  [Eq.~({\ref{cc-1})] 
for the electric density/current are obtained as 
\beqa
K^{\rm cc}_{00} &=& q^{2}K, \\
K^{\rm cc}_{i0} &=& K^{\rm cc}_{0i}
= q_{i}\omega K, \\
K^{\rm cc}_{ij}
&=& i\omega 
\left\{
\langle D\nu \rangle
\left(\delta_{ij}-\frac{q_{i}q_{j}}{q^{2}}\right)
-i\omega K\frac{q_{i}q_{j}}{q^{2}}
\right\}, 
\eeqa
where 
\beqa
K = \frac{\langle D\nu \bar{Y}\rangle+
2\pi \Gt\langle \nu \rangle 
\langle D\nu \rangle 
}{Y_{\upa}Y_{\doa}+2\pi\Gt\langle Y\nu\rangle}. 
\eeqa
The following properties are seen. 

(i) Gauge invariance \cite{Schrieffer64} and charge conservation are satisfied, 
\beqa
K^{\rm cc}_{\mu\nu}q_{\nu}=0, ~~q_{\mu}K^{\rm cc}_{\mu\nu}=0, 
\eeqa
where $q_{\mu} = (-\omega, \qv)$ is a four wavevector.\cite{note2}

(ii) For $\Gt=0$ 
(without spin-flip scattering), 
we have 
\beqa
K = \frac{\langle D\nu\bar{Y}\rangle}{Y_{\upa}Y_{\doa}} 
= \sum_{\sigma}\frac{D_{\sigma}\nu_{\sigma}}{D_{\sigma}q^{2}-i\omega}.
\label{Gamma2=0_c}
\eeqa
This means that up- and down-spin electrons diffuse independently, 
and there are two independent diffusion modes. 

(iii) For $\Gt \neq 0$, and in the long-wavelength and low-frequency 
limit, 
$\tau_{\rm sf}^{-1} \equiv 2\pi \Gt \langle \nu \rangle /\hbar \gg |Y_{\sigma}|$, 
we have 
\beqa
 K= \frac{\langle \nu \rangle 
 \langle D\nu\rangle}{\langle D\nu\rangle q^{2}-i\omega \langle \nu \rangle }
= \frac{\sigma_{\rm c}/e^{2}}{D_{\rm eff}q^{2}-i\omega}
\eeqa
where 
\beqa
 D_{\rm eff} = \frac{\langle D\nu\rangle}{\langle \nu\rangle} 
= \frac{D_{\upa}\nu_{\upa}+ D_{\doa}\nu_{\doa}}{\nu_{\upa}+\nu_{\doa}}
\eeqa
is the effective diffusion constant, and 
\beqa
\sigma_{\rm c} = e^{2}\langle D\nu\rangle = 
e^{2}\sum_{\sigma}D_{\sigma}\nu_{\sigma}
\eeqa
is the electrical conductivity. 
 There is only one diffusion mode owing to the spin mixing $\Gt$. 
 In the opposite limit, $\tau_{\rm sf}^{-1} \ll |Y_{\sigma}|$, 
we have the behavior (\ref{Gamma2=0_c}).

Finally, the charge density $\rho \equiv \langle j_0(\qv) \rangle_{\omega} $ 
and the current density ${\bm j} \equiv \langle j_i(\qv) \rangle_{\omega}$ 
are given by 
\beqa
\rho &=& -e^{2}K{\rm div}{\bm E}, 
\label{density}
\\
 {\bm j} &=& \sigma_{\rm c}{\bm E} 
 + e^{2}\frac{\langle D^{2}\nu \bar{Y}\rangle + 2\pi \Gt\langle D\nu \rangle^{2}}
 {Y_{\upa}Y_{\doa}+2\pi\Gt\langle Y\nu\rangle}\nabla ({\rm div}{\bm E}), 
\label{current}
\eeqa
where 
${\bm E}(\qv,\omega)$ is a Fourier component of the electric field:
${\bm E}(\qv,\omega) 
= -i\qv\phi^{\rm em}(\qv,\omega)+i\omega{\bm A}^{\rm em}(\qv,\omega)
$ with  ${\rm div}{\bm E} = i\qv \cdot {\bm E}$ and 
$\nabla ({\rm div}{\bm E}) =
i\qv(i\qv \cdot {\bm E})$.

\subsubsection{Spin channel}
 The response functions $K^{\rm sc}_{\mu\nu}(\qv,\omega+i0)$ [Eq.~(\ref{sc-4})]  
for spin density/currents are obtained as 
\beqa
 K^{\rm sc}_{00} &=& 
 q^{2}(K^{\rm s}+\Delta K^{\rm s}), \\
 K^{\rm sc}_{0i} &=& q_{i}\omega (K^{\rm s}+\Delta{K}^{\rm s}), \\
 K^{\rm sc}_{i0} &=& q_{i}\omega K^{\rm s}, \\
 K^{\rm sc}_{ij}
&=& i\omega 
 \left\{
 \langle \sigma D\nu \rangle
 \left( \delta_{ij} - \frac{q_{i}q_{j}}{q^{2}} \right)
 -i\omega K^{\rm s} \frac{q_{i}q_{j}}{q^{2}}
 \right\}, 
\eeqa
with 
\beqa
  K^{\rm s} 
&=& \frac{\langle \sigma D\nu \bar{Y}\rangle + 2\pi \Gt \langle \nu \rangle 
     \langle \sigma D\nu \rangle}{Y_{\upa}Y_{\doa} + 2\pi\Gt\langle Y\nu\rangle}, 
\\
  K^{\rm s} + \Delta K^{\rm s} 
&=& \frac{\langle \sigma D\nu \bar{Y}\rangle + 2\pi \Gt \langle \sigma \nu \rangle 
          \langle D\nu \rangle }{Y_{\upa} Y_{\doa} + 2\pi \Gt \langle Y \nu \rangle} . 
\eeqa
The difference 
\beqa
 \Delta K^{\rm s} 
&=& 
 2\pi \Gt\frac{\langle \sigma \nu \rangle 
 \langle D\nu \rangle - \langle \nu \rangle 
 \langle \sigma D\nu\rangle}
 {Y_{\upa}Y_{\doa}+2\pi\Gt\langle Y\nu\rangle}
\nonumber\\
&=& 2\pi\Gt
 (\sigma_{\rm c} \nu_{-} - \sigma_{\rm s}\nu_{+})/Ze^2
\nonumber\\
&=& 2\pi\Gt \nu_{+}\sigma_{\rm c}(P_{\nu}-P_{j})/Ze^2 
\label{Delta_Ks}
\eeqa
arises if $\Gt \neq 0$ (and $P_{\nu}\neq P_{j}$). 
 In Eq.~(\ref{Delta_Ks}),  
\beqa
\sigma_{\rm s} = e^{2}\langle \sigma D \nu\rangle=
e^{2}\sum_{\sigma}\sigma D_{\sigma}\nu_{\sigma}
\eeqa
is the \lq spin conductivity', and $P_{\nu} = \nu_{-}/\nu_{+}$ 
and $P_{j} = \sigma_{\rm s}/\sigma_{\rm c}$ represent 
spin asymmetry in the density of states and in current density, respectively,  
which are different in general. 
The following properties are seen. 

(i) Gauge invariance is satisfied, 
\beqa
 K^{\rm sc}_{\mu\nu} q_{\nu}=0, 
\label{gauge-inv-sc}
\eeqa
but spin conservation is not,
\beqa
q_{\mu}K^{\rm sc}_{\mu\nu} = -\left( q^2 \delta_{\nu 0} + \omega q_i \delta_{\nu i} 
\right)\omega\Delta {K}^{\rm s} \neq 0 , 
\label{spin-con-sc}
\eeqa
if $\Gt \neq 0$, 
where $i$ is a space component.\cite{subscripts2}

(ii) Depending on the relative magnitude of $\tau_{\rm sf}^{-1}$ and $|Y_{\sigma}|$, 
there are two regimes similarly to the charge channel. 
 More interestingly, however, for $\tau_{\rm sf}^{-1} \gg |Y_{\sigma}|$, 
the magnitudes of $\rho_{\rm s}$ and ${\bm j}_{\rm s}$ can be independent, 
governed, respectively, by asymmetry in density of states and by asymmetry in conductivity; 
$\rho_{\rm s} \propto P_\nu \sigma_{\rm c}$ and ${\bm j}_{\rm s} \propto \sigma_{\rm s}$.

Finally, the spin density $\rho_{\rm s} \equiv \langle j_{{\rm s},0}(\qv) \rangle_{\omega}$ 
and the spin-current density ${\bm j}_{\rm s} \equiv \langle j_{{\rm s},i}(\qv) \rangle_{\omega}$ 
are given by 
\beqa
\label{spin-density}
 \rho_{\rm s} &=& -e^{2}({K}^{\rm s}
 +\Delta{K}^{\rm s}){\rm div}{\bm E}, 
\\
\label{spin-current}
 {\bm j}_{\rm s} &=& \sigma_{\rm s}{\bm E} 
 + e^{2}\frac{
 \langle \sigma D^{2}\nu \bar{Y}\rangle
 +2\pi\Gt\langle D\nu \rangle
 \langle \sigma D\nu \rangle
 }{Y_{\upa}Y_{\doa}+2\pi\Gt
\langle Y\nu\rangle}\nabla ({\rm div}{\bm E}) . 
\nonumber\\
\eeqa

\subsubsection{Spin-resolved channel}

 From Eqs.~(\ref{density}), (\ref{current}), 
(\ref{spin-density}) and (\ref{spin-current}), 
we obtain the \lq\lq spin-resolved'' density and current, 
\beqa
&&\rho_{\sigma} = -e^{2}K_{\sigma}{\rm div}{\bm E}, 
\label{rho_sigma} \\
&&{\bm j}_{\sigma} = 
\sigma_{\sigma}{\bm E} + e^{2}D_{\sigma}K_{\sigma}
\nabla({\rm div}{\bm E}), 
\label{j_sigma}
\eeqa
where
\beqa
K_{\sigma} = \frac{
D_{\sigma}Y_{\bars}
+
2\pi\Gt\langle{D\nu}\rangle
}{Y_{\upa}Y_{\doa}+2\pi\Gt\langle Y \nu \rangle}\nu_{\sigma}. 
\label{K_sigma}
\eeqa
 From Eqs.~(\ref{rho_sigma}) and (\ref{j_sigma}), we may derive 
\beqa
{\bm j}_{\sigma} = \sigma_{\sigma}{\bm E} - D_{\sigma} \nabla \rho_{\sigma}
\eeqa
where
\beqa
\sigma_{\sigma}=e^{2}D_{\sigma}\nu_{\sigma}  
\label{sigma}
\eeqa
is the \lq\lq spin-resolved'' conductivity. 
 Further discussion will be given in Sec.~VI.

\section{Spin and Charge Transport in time-dependent spin texture}
In the previous section, 
we studied spin and charge transport 
in a ferromagnetic conductor in its uniformly magnetized state. 
In the second part of this paper, which consists of Sec.~IV and Sec.~V, 
we consider a more general case 
in which the magnetization varies in space and time. 
 This magnetic texture and dynamics induce density change and current 
even if $A_\mu^{\rm em}$ is absent, which are calculated in this paper 
in the first order in both spatial gradient and time derivative.

\subsection{Transformation to local spin frame}
To treat the effects of space- and time-dependent magnetization, 
we introduce a local spin frame where  
the spin quantization axis of $s$-electrons 
is taken to be the $d$-spin direction $\nv(x)$ at each space-time point. \cite{{KMP77},{Volovik87},{TF94}}  
 The original spinor $c$ is then transformed to a spinor $a$ 
in the new frame (rotated frame) as $c = Ua$, 
where $U$ is a 2 $\times$ 2 unitary matrix 
satisfying $c^{\dagger} (\nv\cdot{\bm \sigma}) \, c = a^{\dagger}\sigma^{z} a$. 
 It is convenient to take $U= {\bm m}\cdot{\bm \sigma}$ with  
\beqa
 {\bm m} = \left(\sin\frac{\theta}{2}\cos\phi,~\sin \frac{\theta}{2}\sin\phi,~
\cos\frac{\theta}{2}\right) ,  
\eeqa
where $\theta$ and $\phi$ are ordinary spherical angles parametrizing ${\bm n}$. 
 From space/time derivatives, $\partial_{\mu}c = U(\partial_{\mu}+iA_{\mu})a$, 
there arises an SU(2) gauge field  
\beqa
A_{\mu} = -iU^{\dagger}\partial_{\mu}U = A_{\mu}^{\alpha}
\sigma^{\alpha}. 
\eeqa
 This is an effective gauge field, which represents space/time variations of magnetization. 
 The Lagrangian in the rotated frame is then given by 
$L = \tilde{L}_{\rm el}-H_{\rm e-A}$, 
\beqa
&&\tilde{L}_{\rm el} = \int d\rv~ a^{\dagger}
\left[i\hbar
\frac{\partial}{\partial t}+
\frac{\hbar^2}{2m}\nabla^{2} + \vare_{\rm F} - \tilde{V}_{\rm imp} + M \sigma^{z}
\right]a, \nonumber\\
&&
\label{Lagrangian-e}\\
&& H_{\rm e-A} 
= -\frac{\hbar}{e} \int d\rv~\tilde{j}^\alpha_\mu A^{\alpha}_\mu 
+\frac{\hbar^{2}}{2m} \int d\rv~A^{\alpha}_{i}A^{\alpha}_{i}
a^{\dagger}a ,
\label{Hamiltonian-e-A} 
\eeqa
where 
$\tilde j^\alpha_\mu = (\tilde{\rho}^\alpha, \, \tilde {\bm j}^\alpha)$ 
is a four current representing spin and spin-current densities (\lq\lq paramagnetic'' component) 
in the rotated frame, 
\beqa
&& \tilde{\rho}^\alpha
= -e a^{\dagger} \sigma^{\alpha} a  \ \ (\ = \, \tilde{j}^\alpha_0), \\
&& \tilde{\bm j}^\alpha = -e\frac{\hbar}{2mi} \, 
   a^{\dagger} \sigma^{\alpha} \overset{\leftrightarrow}{\nabla} a. 
\eeqa 
The spin part of the impurity potential $\tilde{V}_{\rm imp}$ is expressed as 
$S^{\alpha}_{j}(c^{\dagger}\sigma^{\alpha}c) 
=
\tilde{S}^{\alpha}_{j}(t)(a^{\dagger}\sigma^{\alpha}a)$, 
where 
$\tilde{S}_{j}^{\alpha}(t) = {\cal R}^{\alpha\beta}({\bm R}_{j}',t)S^{\beta}_{j}$ is the impurity spin in the rotated frame 
{\cite{KS07}} 
with 
\beqa
{\cal R}^{\alpha\beta}= 2m^{\alpha}m^{\beta}-\delta^{\alpha\beta}
\eeqa
being a $3\times 3$ orthogonal matrix representing the same rotation as $U$. 
 Hereafter, the anisotropy axis of impurity spins is defined in reference to the rotated frame  
\begin{equation}
  \overline{\tilde S_i^\alpha \tilde S_j^\beta} 
= \delta_{ij} \delta^{\alpha\beta} \times \left\{ \begin{array}{cc} 
  \overline{S_\perp^2} & (\alpha, \beta = x,y) \\ 
  \overline{S_z^2}     & (\alpha, \beta = z) 
  \end{array} \right. .
\end{equation}

\subsection{Effective U(1) gauge field}

There is some arbitrariness in the choice of the rotated frame;  
one could take $c= U'a'$ with $U'= Ue^{-i\sigma^{z}\chi/2}$, 
where $\chi$ is an arbitrary function of $x$. 
 This arbitrariness is a gauge degree of freedom in the sense that physical quantities  
should not depend on it. 
 It is in fact expressed as the gauge transformation on $a$ and $A_{\mu}$,  
\beqa
 a'&=& e^{-i\sigma^{z}\chi/2}a, 
\label{gauge-trans-a} \\
 A'_{\mu} &=& -i(U')^{\dagger}\partial_{\mu}U' 
\nonumber\\
&=& e^{i\sigma^{z}\chi/2}A_{\mu}e^{-i\sigma^{z}\chi/2}
-\sigma^{z}\partial_{\mu}\chi/2 , 
\label{gauge-trans-perp1}
\eeqa
or, in componentwise, 
\beqa
\label{gauge-trans-perp2}
 A'^{x}_{\mu}+iA'^{y}_{\mu}
&=& e^{-i\chi}\left(A^{x}_{\mu}+iA^{y}_{\mu}\right) ,
\\
\label{gauge-trans-z}
A'^{z}_{\mu}&=& A^{z}_{\mu} - \partial_{\mu}\chi/2 . 
\eeqa
 Note that its $z$ component $A^{z}_{\mu}$ transforms like a gauge potential in ordinary electromagnetism, 
hence can be regarded as a U(1) gauge field. 
 In the following, when we refer to gauge transformation, it means Eqs.~(\ref{gauge-trans-a})-(\ref{gauge-trans-z}). 
 In the next subsection, we study spin and charge transport 
driven by magnetization dynamics as a linear response to this effective gauge field $A^{z}_{\mu}$.

 Generally, one can do a gradient expansion in terms of $A^\alpha_{\mu}$. 
 The expansion parameter is 
$q v_{{\rm F}\sigma}\tau_{\sigma}$ and $\omega \tau_{\sigma}$ 
(for $A^{z}_{\mu}$),\cite{SK09}  
where $q^{-1}$ and $\omega$ are characteristic length and frequency, respectively, of the magnetic texture. 
 In this work, we consider only the lowest nontrivial order in the expansion 
by assuming $q v_{{\rm F}\sigma}\tau_{\sigma} \ll 1$ and $\omega \tau_{\sigma} \ll 1$.
 This condition coincides with the condition, 
$|X_{\sigma}|= |D_{\sigma}q^{2}-i\omega|\tau_{\sigma}\ll 1$, declared below Eq.~(\ref{eq:X}). 
 In typical experiments with Permalloy
($v_{{\rm F}\sigma} \sim 10^5$m/s, $\tau_{\sigma} \sim 10^{-14}$s)
{\cite{Bass-Pratt07}}, 
$q^{-1} \sim$ 100 nm, $\omega \sim$ 100 MHz 
{\cite{YBKXNTE09}}, 
we have $D_\sigma q^2 \tau \sim 10^{-4}$ and $\omega \tau_\sigma \sim 10^{-6}$, 
and the above conditions are satisfied quite well.

\subsection{Linear response to  $A^{\rm em}_{\mu}$ and $A^{z}_{\mu}$.} 

Let us examine the density/current response to the two gauge fields, $A^{\rm em}_{\mu}$ and $A^{z}_{\mu}$. 
 Spin density and currents considered here are the ones whose spin is projected on ${\bm n}$  
(or $\hat z$ in the rotated frame), 
{\it i.e.}, $\rho_{\rm s} = \tilde\rho^{\, z}$ and 
$\tilde{{\bm j}}_{\rm s} = \tilde{{\bm j}}^z$.  
 The total current densities contain the gauge fields, 
\beqa
  j_{\mu} &=& (\, \rho \, , \, 
 \tilde {\bm j} +(e \rho {\bm A}^{\rm em} + \hbar \tilde\rho^\alpha {\bm A}^\alpha)/m ) ,
\\ 
 j_{{\rm s},\mu} &=& (\, \rho_{\rm s} \, , \, 
 \tilde {\bm j}_{\rm s} + ( e \rho_{\rm s} {\bm A}^{\rm em}
   + \hbar \rho {\bm A}^z)/m ) , \ \  
\eeqa
for charge and spin channels, where $\rho = -e a^\dagger a$ and 
$\tilde {\bm j}=(-e\hbar/2mi) \, a^\dagger \overset{\leftrightarrow}{\nabla}a$.  
 By generalizing Eqs.~(\ref{current-1}) and (\ref{spin-current-1}), we may write 
\beqa
\label{cc-cs}
\langle j_{\mu}(\qv) \rangle_{\omega} 
&=& e^2 \tilde{K}^{\rm cc}_{\mu\nu}A^{\rm em}_{\nu} 
+ e\hbar \tilde{K}^{\rm cs}_{\mu\nu}A^{z}_{\nu}  , 
\\
\langle j_{{\rm s},\mu}(\qv) \rangle_{\omega} 
&=& e^2  \tilde{K}^{\rm sc}_{\mu\nu}A^{\rm em}_{\nu} 
+ e\hbar \tilde{K}^{\rm ss}_{\mu\nu}A^{z}_{\nu} . 
\label{sc-ss}
\eeqa
 The response functions, 
$\tilde{K}^{\rm cc}_{\mu\nu}$ and $\tilde{K}^{\rm sc}_{\mu\nu}$, 
are obtained from Eqs.~(\ref{cc-1}) and (\ref{sc-1}) by replacing  
the electron operators in the original frame, $c$ $(c^{\dagger})$, by those in the rotated frame, $a$ $(a^{\dagger})$, 
and are already calculated as $K^{\rm cc}_{\mu\nu}$ and $K^{\rm sc}_{\mu\nu}$ in Sec. III. 
 Thus the response to $A^{\rm em}_{\mu}$ in Eqs.~(\ref{cc-cs}) and (\ref{sc-ss}) 
exactly follows the results there. 

 Let us then focus on the response to $A^{z}_{\mu}$, in  particular, on $\tilde{K}^{\rm cs}_{\mu\nu}$. 
($\tilde{K}^{\rm ss}_{\mu\nu}$ will be presented in Appendix D.)  
 From the definition (linear-response formula), one can show that 
the Onsager's reciprocity relations hold,  
\beqa
\tilde{K}^{\rm cs}_{\mu\nu}(\qv,i\omega_{\lambda}) = 
\tilde{K}^{\rm sc}_{\nu\mu}(-\qv,-i\omega_{\lambda}) ,
\label{Onsager-1}
\eeqa
or 
\beqa
\tilde{K}^{\rm cs}_{\mu\nu}(\qv,\omega+i0) = 
\tilde{K}^{\rm sc}_{\nu\mu}(-\qv,-\omega-i0) .
\label{Onsager-2}
\eeqa
 From this, we see that 
\beqa
q_{\mu}\tilde{K}^{\rm cs}_{\mu \nu} 
= \tilde{K}^{\rm sc}_{\nu\mu}q_{\mu} = 0 ,
\eeqa
namely, the charge conservation is satisfied also in the response to $A^{z}_{\mu}$. 
 On the other hand, if $\Gt \ne 0$, spin is not conserved, 
$q_{\nu}\tilde{K}^{\rm sc}_{\nu\mu} \ne 0$ as seen before. 
 This fact, combined with Eq.~(\ref{Onsager-2}), implies that $\tilde{K}^{\rm cs}_{\mu\nu}$ 
is not gauge invariant,   
\beqa
 \tilde{K}^{\rm cs}_{\mu\nu} q_{\nu}  
= q_{\nu }\tilde{K}^{\rm sc}_{\nu\mu} 
\neq 0,
\eeqa
if $\Gt \neq 0$. 
 The gauge non-invariant terms in Eq.~(\ref{cc-cs}) may be extracted as\cite{subscripts2}  
\beqa
  j_{\mu}'(\qv,\omega )
= e\hbar \Delta K^{\rm s} \{ q^2 \delta_{\mu 0} + q_i \omega \delta_{\mu i} \} A^{z}_{\qv,0} . 
\label{gauge-non-inv-term}
\eeqa

 To summarize, the calculation based on the gauge field $A^{z}_{\mu}$  
fails to respect gauge invariance in the presence of spin-flip scattering. 
 Stated more explicitly, the density and current calculated {\it as a linear response to} $A^z_{\mu}$ 
are not gauge invariant.\cite{note3}

\section{Careful treatment of spin relaxation effects}
\subsection{Restoration of gauge invariance}

The lack of gauge invariance encountered in Sec.~IV-C is due to an oversight 
of some contributions. 
 We recall that the quenched magnetic impurities in the original frame 
become time-dependent in the rotated frame,  
$\tilde{S}_{j}(t) = {\cal R}^{\alpha\beta}({\bm R}'_{j},t)S^{\beta}_{j}$. 
Therefore, we should treat the spin part of the impurity potential 
\beq
\label{p-Hamiltonian}
 H_{\rm s} = u_{\rm s}\sum_{j}
 \int d\rv \tilde{{\bm S}}_{j}(t)\delta(\rv-{\bm R}'_{j}) \cdot (a^{\dagger}{\bm \sigma}a)_{x}  
\eeq
as a time-dependent perturbation.
 The same situation was met in the calculation of Gilbert damping. \cite{KS07}  

Since the first-order (linear) response vanishes, 
$\overline{\tilde{S}^{\alpha}_{j}(t)}=0$, 
let us consider the second-order (nonlinear) response, 
\beqa
\label{extra-current}
&& \Delta j_{\mu} (\qv,\omega) \nonumber\\
&=& -en_{\rm s}u_{\rm s}^{2}\int_{-\infty}^{\infty}\frac{d\omega'}{2\pi}
  \chi^{\alpha\beta}_{\mu}(\qv;\omega,\omega')
\overline{\big[
\tilde{S}^{\alpha}(\omega-\omega')
\tilde{S}^{\beta}(\omega')
\big]}_{\qv}\nonumber\\
\eeqa
where 
$\tilde{S}^{\alpha}_{\pv}(\omega)$ is the Fourier component of 
$\sum_j \tilde{S}^{\alpha}_{j}(t) \, \delta ({\bm r} - {\bm R}_j')$, 
and $\chi_{\mu}^{\alpha\beta}$
is the nonlinear response function. \cite{KS07} 
 To calculate it, it is simpler to use the path-ordered Green's function.\cite{{Rammer86}}
 The contribution represented in Fig.~\ref{nonlinear_1} are given by  
\beqa
&& \chi_{\mu}^{\alpha\beta}(\qv;\omega,\omega')
=\sum_{\kv,\kv'}\int_{-\infty}^{\infty}\frac{d\vare}{2\pi i}
\nonumber\\
&&\times
 {\rm tr}[(v_{\mu}+\Lambda_{\mu})G_{\kv_{+}}(\vare_{+})\sigma^{\alpha}
 G_{\kv'}(\vare+\omega')\sigma^{\beta}
 G_{\kv_{-}}(\vare)]^{<} 
\nonumber\\
\label{chi-ab-mu-1}
\eeqa
where $\vare_{+}=\vare+\omega$. 
 The Green's function 
$G_{\kv}(\vare)$ now stands for a path-ordered one, whose 
lesser component is given by 
\beqa
G^{<}_{\kv}(\vare) = f(\vare)(G^{\rm A}_{\kv}(\vare)
-G^{\rm R}_{\kv}(\vare)) , 
\eeqa 
with 
$f(\vare)$ being the Fermi distribution function.   
 In Eq.~(\ref{chi-ab-mu-1}), we adopt a matrix notaion, 
$(G)_{\sigma,\sigma'} = G_{\sigma}\delta_{\sigma\sigma'}$,  
$(\Lambda_{\mu})_{\sigma,\sigma'} = \Lambda_{\mu}^{\sigma} \delta_{\sigma\sigma'}$ 
with $\Lambda_{\mu}^{\sigma}$ given by Eq.~(\ref{Lambda-s-nu}), and \lq tr' means trace in spin space.

We expand $\chi^{\alpha\beta}_{\mu}(\qv;\omega,\omega')$ 
with respect to $\omega$ and $\omega'$ as 
\beqa
\chi^{\alpha\beta}_{\mu}(\qv;\omega,\omega')
 = A^{\alpha\beta}_{\mu} - i\omega B^{\alpha\beta}_{\mu} - i\omega'C^{\alpha\beta}_{\mu} 
  + \cdots 
\label{chi-ab-mu2}
\eeqa
where $A^{\alpha\beta}_{\mu}$, $B^{\alpha\beta}_{\mu}$ and $C^{\alpha\beta}_{\mu}$ are the expansion coefficients. 
 Substituting Eq.~(\ref{chi-ab-mu2}) into Eq.~(\ref{extra-current}), 
we have 
\beqa
  \Delta j_{\mu}(\qv,\omega) 
&=& -en_{\rm s} u_{\rm s}^2 \left[
  B^{\alpha\beta}_{\mu} \overline{ \partial_t (\tilde{S}^{\alpha} \tilde{S}^{\beta})} 
+ C^{\alpha\beta}_{\mu} \, \overline{\tilde{S}^{\alpha} \partial_t \tilde{S}^{\beta}} \, 
  \right]_{\qv,\omega} 
\nonumber \\
\label{extra-current-1}
\eeqa
where $\tilde {\bm S} = \tilde {\bm S}(t)$ is time dependent. 
 (We have dropped a term containing $A^{\alpha\beta}_{\mu}$, which does not reflect the time dependence  
of $\tilde {\bm S}(t)$.)  
 From
\beqa
 \overline{\tilde{S}^{\alpha} \partial_t \tilde{S}^{\beta}} 
= (\overline{S^{2}_{\perp}}\delta^{\alpha\gamma}_{\perp}
  + \overline{S^{2}_{z}} \, \delta^{\alpha z}\delta^{\gamma z}) 
  ({\cal R} \partial_t {\cal R})^{\gamma\beta} , 
\eeqa
where 
$\delta^{\alpha\beta}_{\perp}
=\delta^{\alpha\beta} - \delta^{\alpha z} \delta^{\beta z}$, 
and the relation\cite{KS07}  
\beqa
({\cal R} \partial_{\mu} {\cal R})^{\alpha\beta} = 2 \vare^{\alpha\beta\gamma} A^{\gamma}_{\mu} ,  
\eeqa
we see that Eq.~(\ref{extra-current-1}) describes a response to $A^\gamma_0$. 
 The coefficients are calculated as \cite{subscripts2} (see Appendix B) 
\beqa
 B^{\alpha\beta}_{\mu} 
&=& - \frac{1}{2} \, C^{\alpha\beta}_{\mu} , 
\nonumber \\
&=&  \pi \nu_\uparrow \nu_\downarrow 
      \frac{ \langle \sigma Y \rangle \, \delta_{\mu 0} + iq_i \langle \sigma D \bar Y \rangle \, \delta_{\mu i}}
           {Y_{\upa}Y_{\doa}+2\pi\Gt\langle Y\nu\rangle} 
      \, \vare^{\alpha\beta}  \,  
\label{C-ab-mu}
\eeqa
where 
$\vare^{\alpha\beta} = \vare^{\alpha\beta z}$, 
and we have dropped unimportant terms proportional to 
$\delta^{\alpha\beta}_\perp$ or $\delta^{\alpha z}\delta^{\gamma z}$.
\begin{figure}
\includegraphics[scale=0.27]{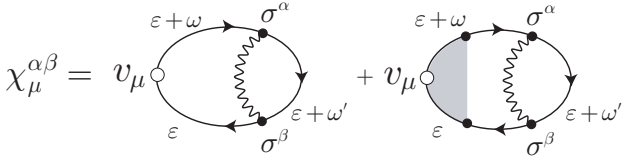}
\caption{Diagrammatic expression of $\chi^{\alpha\beta}_{\mu}$. 
The wavy line represents scattering from impurity spins, 
which are time-dependent in the rotated frame. 
The shaded part represents the vertex function $\Lambda^{\sigma}_{\mu}$. 
} 
\label{nonlinear_1}
\end{figure}
 We thus have 
\beqa
 \Delta j_{\mu} (\qv,\omega) 
&=& e\hbar \Delta \tilde{K}^{\rm cs}_{\mu\nu} A^{z}_{\qv,\nu}, 
\label{gauge-restoration-term}
\eeqa
with\cite{subscripts2}
\beqa
 \Delta \tilde{K}^{\rm cs}_{\mu\nu} 
&=& - \Delta K^{\rm s} \{ q^2 \delta_{\mu 0} + q_i \omega \delta_{\mu i} \}
\delta_{\nu 0}. 
\label{delta_K}
\eeqa
 This new contribution cancels the gauge-dependent terms, Eq.~(\ref{gauge-non-inv-term}), 
and restores the gauge invariance,  
\beqa
(\tilde{K}^{\rm cs}_{\mu\nu}+\Delta \tilde{K}^{\rm cs}_{\mu\nu})q_{\nu} 
= 0 .
\eeqa
 Note that it 
does not affect the charge conservation 
since $q_\mu \Delta \tilde{K}^{\rm cs}_{\mu\nu} = 0$, 
nor the spin non-conservation ($q_{\mu}\tilde{K}^{\rm sc}_{\mu\nu}\neq 0$)   
since it does not contribute to $\tilde{K}^{\rm sc}_{\mu\nu}$.

 The gauge-invariant result for the charge density 
$\rho^{{\rm smf} \, (1)} (\qv,\omega)$ 
and current density ${\bm j}^{{\rm smf} \, (1)}(\qv,\omega)$ induced by magnetization dynamics is summarized as 
\beqa
\label{Density-Az-2}
 \rho^{\rm smf \, (1)} &=& -e^{2} K^{\rm s}{\rm div}{\bm E}_{\rm s}^{0}, \\
\label{Current-Az-2}
 {\bm j}^{\rm smf \, (1)} &=& \sigma_{\rm s}{\bm E}_{\rm s}^{0}
\nonumber\\
&+& e^{2} \frac{\langle\sigma D^{2}\nu \bar{Y}\rangle + 2\pi\Gt\langle D\nu \rangle\langle\sigma D\nu \rangle}
   {Y_{\upa}Y_{\doa}+2\pi\Gt\langle Y\nu\rangle}\nabla
   ({\rm div}{\bm E}_{\rm s}^{0}). 
\nonumber\\
\eeqa
 The first term on the right-hand side of Eq.~(\ref{Current-Az-2}) has the form of Eq.~(\ref{intro-charge-current}), and 
implies the existence of spin-dependent motive force described by the effective \lq electric' field ${\bm E}_{\rm s}^{0}$. 
 The second term of Eq.~(\ref{Current-Az-2}) 
represents a diffusion current arising from charge imbalance 
induced by ${\bm E}_{\rm s}^{0}$, as made clear in Sec.~VI. 
 This term implies the existence of nonlocal spin-transfer torque as the reciprocal effect, 
whose study will be left to the future.

\subsection{Dissipative correction}

\begin{figure}
\includegraphics[scale=0.22]{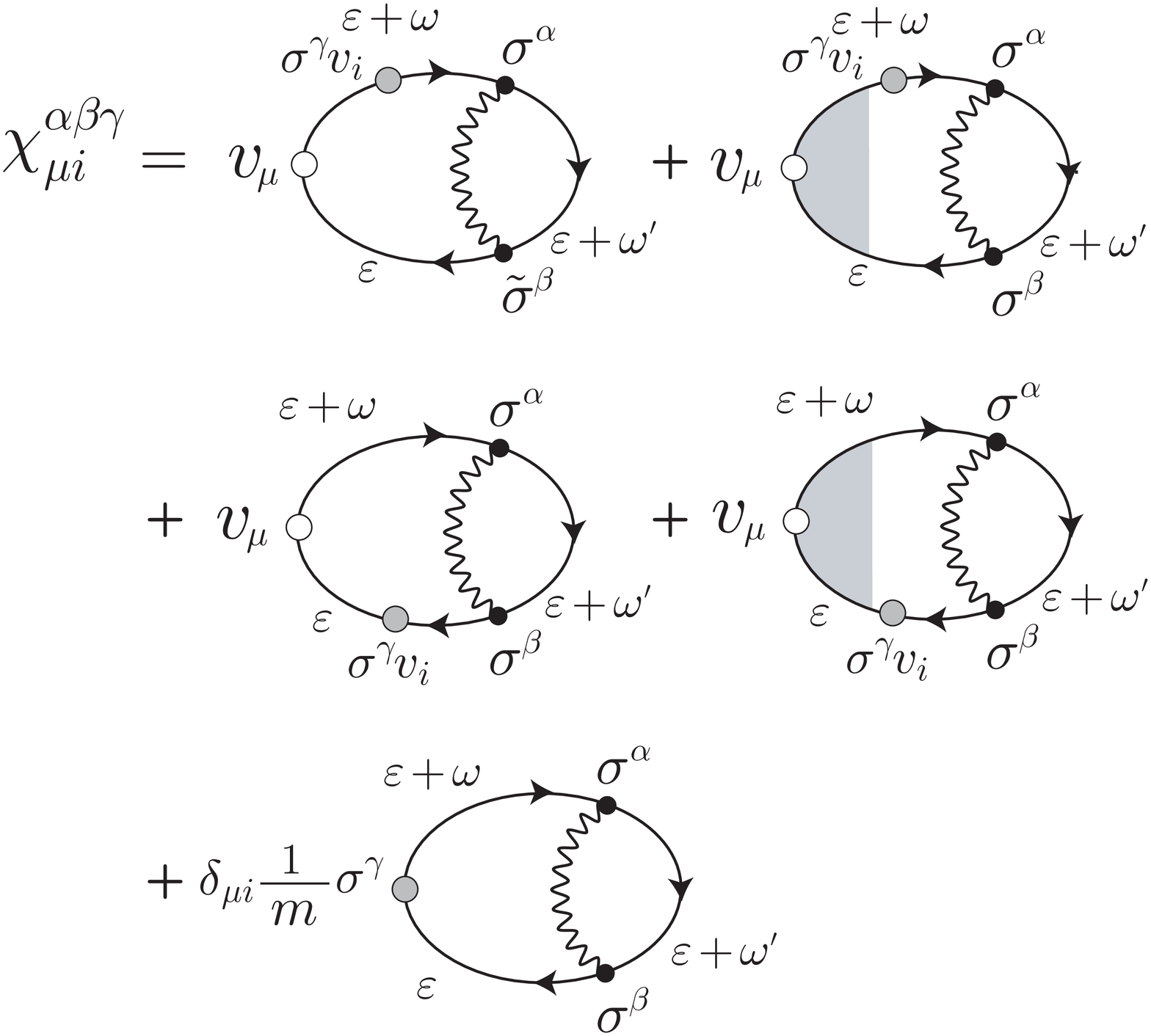}
\caption{Diagrammatic expression of $\chi^{\alpha\beta\gamma}_{\mu i}$. 
The gray circle represents the interaction with $A_{\mu}^\gamma$.
} 
\label{nonlinear_2}
\end{figure}

 It is important to note that 
there is one more contribution within the same order in gradient expansion. 
 It is essentially given by Eq.~(\ref{extra-current}), 
but with one more factor of $A_\mu^\alpha$. 
 The response function, 
denoted by 
$\chi^{\alpha\beta\gamma}_{\mu i}$, is obtained from Eq.~(\ref{chi-ab-mu-1}) 
by further extracting $A_\mu^\alpha$ via Eq.~(\ref{Hamiltonian-e-A}). 
 These are expressed as [Fig. \ref{nonlinear_2}] 
\beqa
\label{j-smf-2}
 j^{{\rm smf} \, (2)}_{\mu}(\qv,\omega)
&=& -e\hbar n_{\rm s} u_{\rm s}^{2} 
     \sum_{\qv'}\int_{-\infty}^{\infty}\frac{d\omega'}{2\pi}
     \chi^{\alpha\beta\gamma}_{\mu i}({\bm q};\omega,\omega')
\nonumber\\
&\times& 
 \overline{\big[\tilde{S}^{\alpha}(\omega-\omega')\tilde{S}^{\beta}(\omega')\big]}_{\qv-\qv'}
A^{\gamma}_{\qv',i}, 
\eeqa
where 
\beqa
&& \chi^{\alpha\beta\gamma}_{\mu i}({\bm q};\omega,\omega') 
= \sum_{\kv,\kv'} \int_{-\infty}^{\infty} \frac{d\vare}{2\pi i} \, 
 {\rm tr} \Big[ (v_{\mu}+\Lambda_{\mu}) 
\nonumber\\
&\times& \{ v_i^+
 G_{\kv+}(\vare_{+}) \sigma^{\gamma}      G_{\kv+}(\vare_{+}) \sigma^{\alpha} 
 G_{\kv'}(\vare +\omega') \sigma^{\beta}  G_{\kv-}(\vare)
\nonumber\\
&&+ v_i^- G_{\kv+}(\vare_{+} )\sigma^{\alpha} G_{\kv'}(\vare +\omega') \sigma^{\beta} 
    G_{\kv-}(\vare) \sigma^{\gamma}     G_{\kv-}(\vare) \} 
\nonumber\\
&+& \frac{1}{m} \,  \delta_{\mu i}
 \sigma^\gamma G_{\kv+}(\vare_{+}) 
 \sigma^{\alpha} G_{\kv'}(\vare +\omega') 
 \sigma^{\beta}  G_{\kv-}(\vare)    \Big]^{<} , 
\label{chi-abc-ij}
\eeqa
with $v_i^\pm = (k_i \pm q_i/2)/m$. 
 We have put $\qv' = {\bm 0}$ in Eq.~(\ref{chi-abc-ij}), but retained ${\bm q}$ and $\omega$.  
 Note that the terms with $\gamma=z$ cancel out, and $A_\mu^z$ does not contribute.  
 In the same way as Sec.~V-A, 
we expand $\chi^{\alpha\beta\gamma}_{\mu i}$ 
with respect to $\omega$ and $\omega'$ as 
$\chi^{\alpha\beta\gamma}_{\mu i} 
= A^{\alpha\beta\gamma}_{\mu i} -i\omega B^{\alpha\beta\gamma}_{\mu i} -i\omega' C^{\alpha\beta\gamma}_{\mu i} + \cdots$ 
and focus on the coefficients $B^{\alpha\beta\gamma}_{\mu i}$ and $C^{\alpha\beta\gamma}_{\mu i}$. 
 Deferring the details to Appendix C, we cite the result 
\beqa
 B^{\alpha\beta\gamma}_{\mu i} 
&=& - \frac{1}{2} \, C^{\alpha\beta\gamma}_{\mu i} 
\nonumber\\
&=& (\delta^{\alpha z} \vare^{\beta\gamma}
   -\delta^{\beta z} \vare^{\alpha\gamma})
   \frac{\nu_{+}}{4M} \sum_{\sigma}\sigma (L^{\sigma}_{i\mu})^{\rm RA} ,  \ \  
\eeqa
where $\nu_+ = \nu_{\uparrow}+\nu_{\downarrow}$, and $L^{\sigma}_{i\mu}$'s are given 
by Eqs.~(\ref{LRAi0}) and (\ref{LRAij}). 
 Note the order of the subscripts, $i\mu$.  
 We thus have 
\beqa
  j_\mu^{{\rm smf} \, (2)} (\qv,\omega) = 
 \beta \frac{e\hbar}{\pi} \sum_{\sigma} \sigma (L^{\sigma}_{i\mu})^{\rm RA}
 ({\bm A}^\perp_i \!\cdot\! {\bm A}^\perp_0)_{\qv,\omega} 
\eeqa
where 
${\bm A}^{\perp}_{\mu} = {\bm A}_{\mu} - \hat{z}\, (\hat{z}\!\cdot\! {\bm A}_{\mu})$, and 
\beqa
 \beta = \frac{\pi}{M} n_{\rm s} u_{\rm s}^2 (\overline{S_\perp^2} + \overline{S_z^2})
        (\nu_\uparrow + \nu_\downarrow ) 
\label{beta}
\eeqa
is a measure of spin relaxation. 
 With the relation  
\beqa
 {\bm A}^{\perp}_{i} \!\cdot {\bm A}^{\perp}_{0} 
= \frac{1}{4} \, \dot{\nv} \!\cdot \partial_{i} \nv , 
\eeqa
which is gauge-invariant under (\ref{gauge-trans-perp2}), 
we finally obtain 
\beqa
 \rho^{\rm smf \, (2)}
&=& -e^{2}K_{\rm s}{\rm div}{\bm E}^{\rm dis}_{\rm s} , 
\label{beta-charge-density2}
\\
 {\bm j}^{{\rm smf} \, (2)} &=&
\sigma_{\rm s}{\bm E}^{\rm dis}_{\rm s}\nonumber\\
&+& e^{2} \frac{\langle\sigma D^{2}\nu \bar{Y}\rangle
             + 2\pi\Gt \langle D\nu \rangle\langle\sigma D\nu \rangle}
               {Y_{\upa}Y_{\doa}+2\pi\Gt\langle Y\nu\rangle} 
    \nabla ({\rm div}{\bm E}_{\rm s}^{\rm dis}) , 
\nonumber\\
\label{beta-current-density2}
\eeqa
where ${\bm E}_{\rm s}^{\rm dis}$ is given by Eq.~(\ref{diss-effective-field}). 
 Since ${\bm E}_{\rm s}^{\rm dis}$ contains $\beta$ as a prefactor, 
Eqs.~(\ref{beta-charge-density2}) and (\ref{beta-current-density2}) come from spin-relaxation processes. 
 This $\beta$ is exactly the same as the coefficient of the $\beta$-term of current-induced torque,\cite{KTS06,KS07}  
consistent with the fact that these are reciprocal to each other.\cite{Duine08,Tserkovnyak08}

\section{Results and discussion}

 The results obtained in this paper are summarized as 
\beqa
 \rho &=& - \frac{\langle D \nu \bar Y F \rangle 
          + 2 \pi \Gt \langle \nu \rangle \langle D \nu F \rangle }
         {Y_\uparrow Y_\downarrow + 2 \pi \Gt \langle Y\nu\rangle} ,  
\label{rho_final}
\\
 {\bm j} &=& \sigma_{\rm c} {\bm E} + \sigma_{\rm s} {\bm E}_{\rm s} 
\nonumber \\
&+&  \frac{\langle D^2 \nu \bar Y \nabla F \rangle 
       + 2 \pi \Gt \langle D \nu \rangle \langle D \nu \nabla F  \rangle }
       {Y_\uparrow Y_\downarrow + 2 \pi \Gt \langle Y \nu \rangle} , 
\label{j_final}
\\
 \rho_{\rm s} &=& - \frac{ \langle \sigma D \nu \bar Y F \rangle
     + 2 \pi \Gt \langle \sigma \nu \rangle \langle D \nu F \rangle }
     {Y_\uparrow Y_\downarrow + 2 \pi \Gt \langle Y \nu \rangle} ,  
\label{rhos_final}
\\
{\bm j}_{\rm s} &=& \sigma_{\rm s}{\bm E} + \sigma_{\rm c} {\bm E}_{\rm s} 
\nonumber \\ 
&+& \frac{ \langle \sigma D^2 \nu \bar Y \nabla F \rangle
           + 2 \pi \Gt  \langle \sigma D \nu \rangle \langle D \nu \nabla F \rangle}
           {Y_\uparrow Y_\downarrow + 2 \pi \Gt \langle Y \nu \rangle } , 
\label{js_final} 
\eeqa
where $F_{\rm c} = e^2 {\rm div}{\bm E}$, $F_{\rm s} = e^2 {\rm div}{\bm E}_{\rm s}$, 
and $F_\sigma = F_{\rm c} + \sigma F_{\rm s}$. 
 The notations are as before; for example, 
$\langle D \nu \nabla F \rangle 
 = D_\uparrow \nu_\uparrow \nabla F_\uparrow + D_\downarrow \nu_\downarrow \nabla F_\downarrow$.
 From these relations [or Eqs.~(\ref{j_sigma_final}) and (\ref{E_sigma_final}) below], 
we identify the spin motive field to be 
\beqa
 E_{{\rm s},i} = \frac{\hbar}{2e} 
 \left\{ - {\bm n} \!\cdot\! (\dot {\bm n} \times \partial_i {\bm n}) 
      + \beta (\dot {\bm n} \!\cdot\! \partial_i {\bm n}) \right\} .
\label{Es_final}
\eeqa

 The spin-resolved density and current are given by 
\beqa
  \rho_{\sigma} 
&=& -e^{2}{\rm div}( K_\sigma {\bm E} + K_\sigma' {\bm E}_{\rm s}), 
\label{rho_sigma_final} 
\\
  {\bm j}_\sigma 
&=&  \sigma_\sigma {\bm E}_\sigma - D_\sigma \nabla \rho_\sigma , 
\label{j_sigma_final}
\\
  {\bm E}_\sigma &=&  {\bm E} + \sigma {\bm E}_{\rm s} , 
\label{E_sigma_final}
\eeqa
where ${\bm E}_\sigma$ is the total field felt by spin-$\sigma$ electrons. 
 The coefficient $K_\sigma$ is given by Eq.~(\ref{K_sigma}), and $K_\sigma'$ by 
\beqa
K_\sigma' 
&=& \frac{\sigma D_\sigma Y_{\bar\sigma} + 2 \pi \Gt \langle \sigma D \nu \rangle}
     {Y_\uparrow Y_\downarrow + 2 \pi \Gt \langle Y \nu \rangle} \, \nu_\sigma . 
\label{K'_sigma_final}
\eeqa
 There are two characteristic regimes depending on the relative magnitude of 
$\tau_{\rm sf}^{-1} \equiv 2 \pi \Gt \langle \nu \rangle /\hbar$ and $|Y_{\sigma}|$. 
 For $\tau_{\rm sf}^{-1} \ll |Y_{\sigma}|$, Eq.~(\ref{rho_sigma_final}) becomes 
\beqa
  \rho_{\sigma} 
&\simeq & - \frac{\sigma_\sigma}{D_\sigma q^2 - i\omega} \, {\rm div} {\bm E}_\sigma , 
\eeqa
meaning that the spin-$\sigma$ electrons respond only to ${\bm E}_\sigma$, not to ${\bm E}_{\bar\sigma}$,
and the two spin components ($\uparrow$ and $\downarrow$) behave independently. 
 In particular, the response to  a spin motive field ${\bm E}_{\rm s}$ (set ${\bm E}={\bm 0}$ for simplicity) 
is opposite in sign between $\uparrow$ and $\downarrow$ electrons. 
 In the opposite limit, $\tau_{\rm sf}^{-1} \gg |Y_{\sigma}|$, Eq.~(\ref{rho_sigma_final}) becomes 
\beqa
  \rho_{\sigma} 
&\simeq & - \frac{\nu_\sigma / \langle \nu \rangle}{D_{\rm eff} q^2 - i\omega} \, 
            {\rm div} (\sigma_\uparrow {\bm E}_\uparrow + \sigma_\downarrow {\bm E}_\downarrow ) ,
\eeqa
where $D_{\rm eff} = \langle D \nu \rangle/\langle \nu \rangle$. 
 In this case, the density of spin-$\sigma$ electrons is affected not only by ${\bm E}_\sigma$ 
but also by ${\bm E}_{\bar\sigma}$. 
 This is due to the strong spin mixing;  as an elementary process, 
$\rho_\sigma$ is induced solely by ${\bm E}_\sigma$, not ${\bm E}_{\bar\sigma}$,   
but subsequent spin-flip processes tends to equilibrate $\rho_\uparrow$ and $\rho_\downarrow$. 
 Note that $\uparrow$ electrons and $\downarrow$ electrons respond to ${\bm E}_{\rm s}$ with the same sign. 
(The common sign is determined by that of $\sigma_\uparrow - \sigma_\downarrow$.)

 The above features oppose the picture of two {\it independent} currents, but they are 
actually described within the conventional two-current model. 
{\cite{Mott64, Fert-Campbell68,Son87,Valet-Fert93}}
 This is best demonstrated by the relation  
\beqa
 \frac{\partial}{\partial t} \rho_{\sigma} + {\rm div} {\bm j}_{\sigma} 
= - \left( \frac{\rho_\sigma}{\tau_{{\rm sf}, \sigma}} 
        - \frac{\rho_{\bar\sigma}}{\tau_{{\rm sf}, \bar\sigma}} \right) ,
\label{continuity_sigma_final}
\eeqa
where  
\beqa
 \tau_{{\rm sf}, \sigma}^{-1} = 2\pi \Gt \nu_{\bar\sigma} / \hbar
\eeqa
is the spin-flip rate for spin-$\sigma$ electrons. 
 The right-hand side of Eq.~(\ref{continuity_sigma_final}) represents a coupling between 
$\uparrow$ and $\downarrow$ electrons. 
 In deriving Eq.~(\ref{continuity_sigma_final}), we have used Eqs.~(\ref{rho_sigma_final}), (\ref{j_sigma_final}), 
(\ref{K'_sigma_final}) and (\ref{K_sigma}), and the relations, 
$\langle \sigma K/\nu \rangle = \langle \sigma D \bar Y\rangle /Z$ and 
$\langle \sigma K'/\nu \rangle = \langle D \bar Y\rangle /Z$. 
 Note that $\rho_\sigma$, being given by Eq.~(\ref{rho_sigma_final}), represents a deviation from the equilibrium value.  
 One may define the deviation of chemical potential, $\delta\mu_\sigma$, from equilibrium by 
\beqa
 \rho_\sigma = -e \nu_\sigma \delta \mu_\sigma . 
\eeqa
 Then Eq.~(\ref{continuity_sigma_final}) can be put in a familiar form 
{\cite{Son87,Valet-Fert93} }
\beqa
 \frac{\partial}{\partial t} \rho_{\sigma} + {\rm div} {\bm j}_{\sigma} 
&=& \frac{\sigma_\sigma}{e} \!\cdot\! 
    \frac{ \delta\mu_\sigma - \delta\mu_{\bar\sigma} }{\ell_\sigma^2}  .
\label{continuity_sigma_final_2}
\eeqa
where $\ell_\sigma = \sqrt{D_\sigma \tau_{{\rm sf}, \sigma}}$ is the spin diffusion length for spin-$\sigma$ 
electrons.

 The present work is therefore within the two-current picture. 
 This fact was implicitly used in identifying the spin motive force on the basis of Eq.~(\ref{intro-charge-current}).

\section{Summary}
 In this paper, we have studied spin and charge transport in a conducting 
ferromagnet driven by two kinds of gauge fields, $A^{\rm em}_{\mu}$ 
and $A^{z}_{\mu}$, which act in charge channel and spin channel, respectively. 
 In particular, we have given a microscopic calculation of spin motive force 
by taking spin-relaxation effects into account.

In the first part, we calculated density and current in both spin and charge channels in response 
to the ordinary electromagnetic field $A^{\rm em}_{\mu}$ in a uniformly magnetized state. 
 We observed a crossover from two diffusion modes to a single mode as the spin-flip rate is increased 
(for a fixed frequency/wavenumber of the disturbance), or as the frequency/wavenumber is decreased 
(for a fixed spin-flip rate). 
 However, if expressed in terms of spin-resolved density and current, the so-called two-current model is shown to hold 
irrespective of the strength of spin-flip scattering.

In the second part, we have developed a microscopic theory of spin motive force 
in the framework of gauge-field method. 
 We readily encountered the problem of gauge non-invariance; 
the current calculated as a linear response to $A^{z}_{\mu}$ depends on the gauge (choice of local spin frame). 
 This fact is intimately related to the non-conservation of spin (due to spin-flip scattering) 
by Onsager reciprocity, hence is robust. 
 This theoretical puzzle was resolved by noting the fact that the spin-dependent scattering terms 
(quenched impurity spins) are time-dependent in the rotated frame. 
 By calculating the second-order (nonlinear) response to this time-dependent perturbation, 
we could recover a gauge-invariant result while keeping the spin non-conservation. 
 The dissipative correction to the ordinary spin motive force, 
which is the inverse to the spin-torque $\beta$-term, is also obtained.

{\it Note added:} 
 After submitting the manuscript, we became aware of a closely related work by Kim {\it et al.}\cite{Kim11}

\acknowledgements

The authors would like to thank G. Tatara for discussions,  
and K.-W. Kim for informing us of Ref.~55. 
This work is partially supported by a Grant-in-Aid from Monka-sho, Japan. 


\appendix

\section{Calculation of response functions $K^{\rm cc}_{\mu\nu}$ and $K^{\rm sc}_{\mu\nu}$}

 In this Appendix, we evaluate the electromagntic response functions 
in the ladder approximation shown in Fig.~\ref{vertex_1}(a).   
 From Eqs.~(\ref{cc-1}) and (\ref{sc-1}), they are written as 
\beqa 
K^{\rm cc}_{\mu\nu}(\qv,i\omega_{\lambda}) &=& 
-T\sum_{n,\sigma}L_{\mu\nu}^{\sigma}
(\qv;i\vare_{n}+i\omega_{\lambda},i\vare_{n}), \label{cc-2}\\
K^{\rm sc}_{\mu\nu}(\qv,i\omega_{\lambda}) &=& 
-T\sum_{n,\sigma}\sigma L_{\mu\nu}^{\sigma}
(\qv;i\vare_{n}+i\omega_{\lambda},i\vare_{n}),
\label{sc-2}
\eeqa
with
\beqa
 L_{\mu\nu}^{\sigma}=\Pi^{\sigma}_{\mu\nu}+ \Pi_{\mu0}^{\sigma} \Lambda^{\sigma}_{\nu} , 
\label{eq:L}
\eeqa
\beqa
&&\Pi^{\sigma}_{\mu\nu}
= \sum_{\kv}v_{\mu}v_{\nu}G_{\kv_{+},\sigma}(i\vare_{n}+i\omega_{\lambda})
G_{\kv_{-},\sigma}(i\vare_{n}) , 
\eeqa
where $\vare_{n}=(2n+1)\pi T$ ($n$: integer) is a fermionic Matsubara frequency. 
 The vertex function $\Lambda^{\sigma}_{\nu}$ satisfies [Fig.~\ref{vertex_1}(b)]
\beqa
\Lambda^{\sigma}_{\nu} = \lambda^{\sigma}_{\nu}
+\Gi\Pi_{\sigma}\Lambda^{\sigma}_{\nu}
+\Gt\Pi_{\bars}\Lambda^{\bars}_{\nu}, 
\label{v-2}
\eeqa
where $\Pi_{\sigma}=\Pi^{\sigma}_{00}$,  
and $\lambda^{\sigma}_{\nu} = \Gi\Pi^{\sigma}_{0\nu}
+\Gt\Pi^{\bars}_{0\nu}$
is the lowest-order contribution. 
The equation (\ref{v-2}) is solved as 
\beqa
 \Lambda^{\sigma}_{\nu} 
= \frac{\lambda^{\sigma}_{\nu} 
       - \Pi_{\bars}(\Gi\lambda^{\sigma}_{\nu}
                    - \Gt\lambda^{\bars}_{\nu})}
{ 1 - \Gi (\Pi_\uparrow + \Pi_\downarrow)
+ (\Gi^2-\Gt^2) \Pi_\uparrow \Pi_\downarrow} .
\label{Lambda-s-nu}
\eeqa
 Performing the analytic continuation, 
$i\omega_{\lambda} \to \omega + i0$ 
and retaining terms up to the first order in $\omega$, 
we obtain
\beqa
 K^{\rm cc}_{\mu\nu}(\qv,\omega+i0) 
&=& \nu_{+} \delta_{\mu0} \delta_{\nu0} 
  + \frac{i\omega}{2\pi}\sum_{\sigma} (L_{\mu\nu}^{\sigma})^{\rm RA} , 
\label{cc-4}\\
 K^{\rm sc}_{\mu\nu}(\qv,\omega+i0) 
&=& \nu_{-} \delta_{\mu0} \delta_{\nu0}
  + \frac{i\omega}{2\pi}\sum_{\sigma}\sigma (L_{\mu\nu}^{\sigma})^{\rm RA} , \ \ 
\label{sc-4}
\eeqa
where 
$\nu_{\pm} = \nu_{\uparrow}\pm\nu_{\downarrow}$.  
 The function $(L_{\mu\nu}^{\sigma})^{\rm RA}$ is obtained via the analytic continuation, 
$i(\vare_{n} + \omega_{\lambda}) \to \vare + \omega + i0$ and $i\vare_{n} \to \vare - i0$, 
as indicated by the superscript \lq\lq RA''. 
 We assume $\gamma_{\sigma} \ll \vare_{{\rm F}\sigma}$, and discard   
$(L_{\mu\nu}^{\sigma})^{\rm RR}$ and $(L_{\mu\nu}^{\sigma})^{\rm AA}$ 
as in usual calculations of transport coefficients.  
 The ${\kv}$-integrals are evaluated up to ${\cal O}(|X_{\sigma}|)$ or ${\cal O}(|X_{\sigma}|^0)$ as 
\beqa
(\Pi_{\sigma})^{\rm RA} &=& 
\sum_{\kv}G^{\rm R}_{\kv+,\sigma}(\omega )G^{\rm A}_{\kv-,\sigma}(0)
\nonumber\\ 
&\simeq& 2\pi \nu_{\sigma}\tau_{\sigma}(1-X_{\sigma}), \\
(\Pi^{\sigma}_{i0})^{\rm RA} &=& 
 \sum_{\kv} v_i
G^{\rm R}_{\kv+,\sigma}(\omega )G^{\rm A}_{\kv-,\sigma}(0) 
\nonumber\\
&\simeq& -2\pi iq_{i} D_{\sigma} \nu_{\sigma} \tau_{\sigma}, \\
 (\Pi^{\sigma}_{ij})^{\rm RA}&=& 
\sum_{\kv}v_{i}v_{j}
 G^{\rm R}_{\kv+,\sigma}(\omega )G^{\rm A}_{\kv-,\sigma}(0) 
\nonumber\\
&\simeq&
 2\pi D_{\sigma}\nu_{\sigma}\delta_{ij}. 
\eeqa
where $D_{\sigma}= v_{{\rm F}\sigma}^{2} \tau_{\sigma}/3$, 
$v_{{\rm F}\sigma}=\hbar k_{{\rm F}\sigma}/m$, 
and $X_{\sigma}=Y_{\sigma}\tau_{\sigma}$ with $Y_{\sigma}= D_{\sigma}q^{2}-i\omega$. 
Using these formulas, we obtain
\beqa
&& (\Lambda^{\sigma}_{0})^{\rm RA} = \frac{Y_{\bars}+2\pi\Gt\langle\nu\rangle}
{Y_{\upa}Y_{\doa}+2\pi\Gt\langle Y\nu \rangle} \!\cdot\! \frac{1}{\tau_\sigma} , 
\label{eq:Lambda_0}
\\
&&(\Lambda^{\sigma}_{i})^{\rm RA} = -iq_{i}\frac{D_{\sigma}Y_{\bars}+2\pi\Gt\langle D\nu \rangle}
{Y_{\upa}Y_{\doa}+2\pi\Gt\langle Y\nu \rangle} \!\cdot\! \frac{1}{\tau_\sigma} , 
\label{eq:Lambda_i}
\eeqa
and thus 
\beqa
\label{LRA00}
 (L^{\sigma}_{00})^{\rm RA} 
&=& 2\pi\nu_{\sigma}\frac{Y_{\bars} + 2\pi \Gt
\langle\nu\rangle}{Y_{\upa}Y_{\doa}+2\pi\Gt\langle Y\nu \rangle},\\
\label{LRAi0}
 (L^{\sigma}_{i0})^{\rm RA} &=&
-2\pi iq_{i}\nu_{\sigma}D_{\sigma}\frac{
Y_{\bars}+2\pi \Gt\langle \nu \rangle}
{Y_{\upa}Y_{\doa}+2\pi\Gt\langle Y\nu \rangle},\\
\label{LRA0i}
 (L^{\sigma}_{0i})^{\rm RA} &=&
-2\pi iq_{i}\nu_{\sigma}\frac{D_{\sigma}Y_{\bars}
+2\pi \Gt\langle D\nu \rangle}
{Y_{\upa}Y_{\doa}+2\pi\Gt\langle Y\nu \rangle}, \\
\label{LRAij}
 (L^{\sigma}_{ij})^{\rm RA} 
&=& 2\pi\nu_{\sigma}D_{\sigma}
\left\{ \left( \delta_{ij}-\frac{q_{i}q_{j}}{q^{2}} \right) \right.
\nonumber\\
&&\left.
-i\omega 
\frac{q_{i}q_{j}}{q^{2}}
\frac{Y_{\bars}+2\pi \Gt\langle\nu\rangle}
{Y_{\upa}Y_{\doa}+2\pi\Gt\langle Y\nu \rangle}\right\} . 
\eeqa

\begin{widetext}

\section{Calculation of $C^{\alpha\beta}_{\mu}$}

The nonlinear response function 
$\chi^{\alpha\beta}_{\mu}$ in Eq.~(\ref{chi-ab-mu-1}) 
is written as 
\beqa
&& \chi^{\alpha\beta}_{\mu}(\qv;\omega,\omega')
= \sum_{\sigma,\sigma'} \big[ (\delta^{\alpha\beta}_{\perp} + i\sigma\vare^{\alpha\beta}) \, \delta_{\sigma'\bar\sigma} 
                      + \delta^{\alpha z} \delta^{\beta z} \delta_{\sigma'\sigma} \big]  
  \int_{-\infty}^{\infty}\frac{d\vare}{2\pi i}
  (L^{\sigma}_{0\mu}(\qv;\vare + \omega,\vare)I_{\sigma'}(\vare+\omega'))^{<}  , 
\label{chi-ab-mu-2}
\eeqa
where $L^{\sigma}_{0\mu}$ is given by Eq.~(\ref{eq:L}), and 
$I_{\sigma}(\vare) = \sum_{\kv}G_{\kv\sigma}(\vare)$. 
 Following the Langreth's method,\cite{{Langreth76},{Haug98}}
the lesser component of 
$L^{\sigma}_{0\mu}(\qv;\vare+\omega,\vare) I_{\bars}(\vare+\omega') \equiv LI$ 
is calculated as 
\beqa
 (LI)^< 
&=& f(\vare) (L^{\rm RA}-L^{\rm RR}) I^{\rm R} 
 + f(\vare+\omega') L^{\rm RA} (I^{\rm A}-I^{\rm R}) 
 + f(\vare+\omega) (L^{\rm AA}-L^{\rm RA}) I^{\rm A} . 
\eeqa
 Note that the ordering of Green's functions in $LI$ is $G(\vare+\omega)G(\vare+\omega')G(\vare)$ 
[see Eq.~(\ref{chi-ab-mu-1})].
The superscripts RA, A etc. specify the analytic branch; for example, 
$L^{\rm RA}(\vare + \omega, \vare ) = L(\vare + \omega +i0, \vare -i0)$, 
$I^{\rm A}(\vare )= I(\vare -i0)$, etc. 
 Thus the coefficients in the expansion 
$\chi^{\alpha\beta}_{\mu}  
 = A^{\alpha\beta}_{\mu} - i\omega B^{\alpha\beta}_{\mu} - i\omega'C^{\alpha\beta}_{\mu} + \cdots$
are obtained as  
\beqa
 B^{\alpha\beta}_{\mu} 
&=& 
\frac{1}{2\pi} \sum_{\sigma,\sigma'} 
     \big[ (\delta^{\alpha\beta}_{\perp} + i\sigma\vare^{\alpha\beta}) \, \delta_{\sigma'\bar\sigma} 
          + \delta^{\alpha z} \delta^{\beta z} \delta_{\sigma'\sigma} \big] 
(L^{\sigma}_{0\mu}(\qv;\omega,0))^{\rm RA} I^{\rm A}_{\sigma'}(0)  ,  
\\
 C^{\alpha\beta}_{\mu} 
&=& 
\frac{i}{\pi} \sum_{\sigma,\sigma'} 
     \big[ (\delta^{\alpha\beta}_{\perp} + i\sigma\vare^{\alpha\beta}) \, \delta_{\sigma'\bar\sigma} 
          + \delta^{\alpha z} \delta^{\beta z} \delta_{\sigma'\sigma} \big] 
(L^{\sigma}_{0\mu}(\qv;\omega,0))^{\rm RA}{\rm Im}I^{\rm R}_{\sigma'}(0)  .  
\label{App_A_result}
\eeqa
 We have retained only the lowest-order term in $\gamma_{\sigma}$. 
 Substituting Eqs.~(\ref{LRA00}) and (\ref{LRA0i}) 
together with $I^{\rm R}_{\sigma}(0) = -i\pi \nu_{\sigma}$ 
(whose real part is dropped consistently with the selfenergy)
into Eq.~(\ref{App_A_result}), we obtain Eq.~(\ref{C-ab-mu}).

\section{Calculation of $C^{\alpha\beta\gamma}_{\mu i}$}

 Consider the nonlinear response function 
$\chi^{\alpha\beta\gamma}_{\mu i}$ given by Eq.~(\ref{chi-abc-ij}). 
 As in Appendix B, we take a lesser component, extract the $\omega'$-linear term, 
and retain terms containing both $G^{\rm R}$ and $G^{\rm A}$ to obtain 
$B^{\alpha\beta\gamma}_{\mu i} = - (1/2) C^{\alpha\beta\gamma}_{\mu i}$ and 
\beqa
 C^{\alpha\beta\gamma}_{\mu i} 
&=& \left. 
 i \frac{\partial}{\partial \omega'} \chi^{\alpha\beta\gamma}_{\mu i}(\omega,\omega') \right|_{\omega'=0}
\nonumber\\
&\simeq&  -i \sum_{\kv} 
{\rm tr} \big[(v_{\mu}+\Lambda_{\mu}^{\rm RA}) v_{i}
 \{
  G_{\kv+}^{\rm R} \sigma^{\gamma} G_{\kv+}^{\rm R}
  \sigma^{\alpha} \hat \nu  \, \sigma^{\beta} G_{\kv-}^{\rm A}
 + G_{\kv+}^{\rm R} \sigma^{\alpha} \hat \nu \, 
   \sigma^{\beta} G_{\kv-}^{\rm A} \sigma^{\gamma} G_{\kv-}^{\rm A} \} \big] 
\nonumber \\
&& - \frac{i}{m} \, \delta_{\mu i} \sum_{\kv} 
 {\rm tr}\big[ \sigma^\gamma  
           G_{\kv+}^{\rm R} \sigma^{\alpha} \hat \nu \, \sigma^{\beta} G_{\kv-}^{\rm A} \big] .
\label{eq:B-1}
\eeqa
 Here $(\Lambda_{\mu}^{\rm RA})_{\sigma\sigma'} = (\Lambda_{\mu}^\sigma)^{\rm RA} \delta_{\sigma\sigma'}$ 
is given by Eqs.~(\ref{eq:Lambda_0})-(\ref{eq:Lambda_i}), 
and $\hat \nu = \sum_{{\bm k}'} (G_{\kv'}^{\rm A}-G_{\kv'}^{\rm R})/2\pi i$ 
is a matrix of density of states, $(\hat \nu)_{\sigma\sigma'} = \nu_\sigma \delta_{\sigma\sigma'}$. 
 In Eq.~(\ref{eq:B-1}), all $G$'s are evaluated at $\varepsilon = 0$ except for those in $\Lambda_{\mu}$  
in which ${\bm q}, \omega$ are retained. 
 Equation (\ref{eq:B-1}) is written as 
\beqa
 C^{\alpha\beta\gamma}_{\mu i} 
&=&
 i \sum_{\sigma} \big[ 
 \delta^{\alpha z} (\sigma \delta^{\beta\gamma}_\perp - i \varepsilon^{\beta\gamma}) \nu_{\bar\sigma} 
-\delta^{\beta z} (\sigma \delta^{\alpha\gamma}_\perp - i \varepsilon^{\alpha\gamma}) \nu_\sigma \big] 
 \left\{ M_{\mu i}^\sigma (\qv,\omega) + \bar M_{\mu i}^{\bar \sigma} (\qv,\omega) \right\} , 
\label{B-2}
\eeqa 
where
\beqa
  M^{\sigma}_{\mu i} (\qv,\omega)
&=& Q^{\sigma}_{\mu i}(\qv) + (\Lambda_{\mu }^{\sigma})^{\rm RA} Q^{\sigma}_{0 i} (\qv), 
\\
  \bar{M}^{\sigma}_{\mu i} (\qv,\omega)
&=& \bar{Q}^{\sigma}_{\mu i} (\qv) + (\Lambda_\mu^\sigma)^{\rm RA} \bar{Q}^{\sigma}_{0 i} (\qv), 
\\
 Q^{\sigma}_{\mu i}(\qv) 
&=& \sum_{\kv} v_{\mu} v_{i} \, 
 G_{\kv_{+}, \sigma}^{\rm R} G_{\kv_{-}, \bars}^{\rm R} G_{\kv_{-}, \sigma}^{\rm A} \big|_{\varepsilon =0} 
\ = \  [\bar Q^{\sigma}_{\mu i}(-\qv)]^* ,  
\eeqa
 In the lowest order in $\gamma_\sigma$, we see that 
\beqa
 M^{\sigma}_{\mu i}(\qv,\omega) 
=  \bar{M}^{\sigma}_{\mu i}(\qv,\omega) 
= -\frac{\sigma}{2M} (L^{\sigma}_{i\mu})^{\rm RA} ,
\eeqa
where $(L^{\sigma}_{i\mu})^{\rm RA}$ is given by Eqs.~(\ref{LRAi0}) and (\ref{LRAij}). 
 Noting that 
$M_{\mu i}^\sigma + \bar M_{\mu i}^{\bar \sigma} 
 = -\sum_\sigma \sigma (L^{\sigma}_{i\mu})^{\rm RA}/2M$
is independent of $\sigma$, we obtain the leading term as 
\beqa
 C^{\alpha\beta\gamma}_{\mu i} 
= - (\delta^{\alpha z} \vare^{\beta\gamma} - \delta^{\beta z} \vare^{\alpha\gamma})
\frac{\nu_{+}}{2M} \sum_{\sigma} \sigma (L^{\sigma}_{i\mu})^{\rm RA} . 
\eeqa

\vfill\eject

\end{widetext}


\section{Spin current induced by spin motive force}

The response function 
$\tilde{K}^{\rm ss}_{\mu\nu}$ in Eq.~(\ref{sc-ss}) is evaluated as  
\beqa
 \tilde{K}^{\rm ss}_{\mu\nu} 
&=& \nu_{+} \delta_{\mu 0} \delta_{\nu 0} 
+ \frac{i\omega}{2\pi} \sum_{\sigma} \sigma (L^{\sigma}_{{\rm s},\mu\nu})^{\rm RA}, 
\\
 L^{\sigma}_{{\rm s},\mu\nu} 
&=& \sigma \Pi^{\sigma}_{\mu\nu} + \Pi_{\mu 0}^\sigma \Lambda^{\sigma}_{{\rm s},\nu}  . 
\eeqa
 The spin-current vertex function $\Lambda^{\sigma}_{{\rm s},{\mu}}$, 
which satisfies 
\beqa
 \Lambda^{\sigma}_{{\rm s},\nu} 
= \lambda^{\sigma}_{{\rm s},\nu}
+ \Gi \Pi_{\sigma} \Lambda^{\sigma}_{{\rm s},\nu}
- \Gt \Pi_{\bars} \Lambda^{\bars}_{{\rm s},\nu}, 
\eeqa
with
$\lambda^{\sigma}_{{\rm s},\nu} = \sigma (\Gi \Pi^{\sigma}_{0\nu}
- \Gt \Pi^{\bars}_{0\nu})$,  
is given by 
\beqa
 (\Lambda^{\sigma}_{{\rm s},0})^{\rm RA} 
&=& \frac{\sigma Y_{\bars} + 2\pi \Gt \langle{\sigma\nu}\rangle}
         {Y_{\uparrow} Y_{\downarrow} + 2\pi \Gt \langle Y \nu \rangle}
    \frac{1}{\tau_{\sigma}} , 
\\
 (\Lambda^{\sigma}_{{\rm s},i})^{\rm RA} 
&=& -iq_i \frac{ \sigma D_{\sigma}Y_{\bars} + 2\pi \Gt \langle{\sigma D\nu}\rangle}
               {Y_{\uparrow} Y_{\downarrow} + 2\pi \Gt \langle Y \nu \rangle}
          \frac{1}{\tau_{\sigma}} . 
\eeqa
 Hence, we have
\beqa
\label{app-LRA00}
 (L^{\sigma}_{{\rm s},00})^{\rm RA} 
&=& 2\pi\nu_{\sigma}
 \frac{\sigma Y_{\bars} + 2\pi\Gt \langle \sigma \nu \rangle}
      {Y_{\upa}Y_{\doa} + 2\pi\Gt\langle Y\nu \rangle} , 
\\
\label{app-LRAi0}
(L^{\sigma}_{{\rm s},i0})^{\rm RA} 
&=& -2\pi iq_{i} D_{\sigma} \nu_{\sigma}
 \frac{\sigma Y_{\bars} + 2\pi \Gt \langle \sigma \nu \rangle}
     {Y_{\upa}Y_{\doa} + 2\pi\Gt\langle Y\nu \rangle} , 
\\
\label{app-LRA0i}
 (L^{\sigma}_{{\rm s},0i})^{\rm RA} 
&=& -2\pi iq_{i} \nu_{\sigma}
 \frac{\sigma D_{\sigma}Y_{\bars} + 2\pi \Gt \langle \sigma D \nu \rangle}
      {Y_{\upa}Y_{\doa} + 2\pi\Gt\langle Y\nu \rangle} , 
\\
\label{app-LRAij}
 (L^{\sigma}_{{\rm s},ij})^{\rm RA} 
&=& 2\pi D_{\sigma} \nu_{\sigma} \left\{ \sigma \delta_{ij} - q_i q_j
 \frac{\sigma D_{\sigma}Y_{\bars} + 2\pi\Gt \langle \sigma D\nu\rangle}
      {Y_{\upa}Y_{\doa} + 2\pi \Gt \langle Y\nu \rangle} \right\} . 
\nonumber \\
\eeqa
 Note that $\tilde{K}^{\rm ss}_{\mu\nu}$'s thus obtained do not satisfy 
spin conservation nor gauge invariance, 
$q_\mu \tilde{K}^{\rm ss}_{\mu\nu} = \tilde{K}^{\rm ss}_{\nu\mu} q_\mu \ne 0$, if $\Gt \ne 0$.

\begin{widetext}

 As in Sec.~V, time-dependent magnetic impurities, 
Eq.~(\ref{p-Hamiltonian}), in the rotated frame also induce a spin current 
\beqa
\label{app-extra-spin-current}
 \Delta j_{{\rm s},\mu}(\qv,\omega) 
&=& -e n_{\rm s} u_{\rm s}^{2} \int_{-\infty}^{\infty} \frac{d\omega'}{2\pi} \, 
     \chi^{\alpha\beta}_{{\rm s},\mu}(\qv;\omega,\omega')
 \overline{ \big[ \tilde{S}^{\alpha}(\omega-\omega') \tilde{S}^{\beta}(\omega') \big] }_{\qv} 
\nonumber\\
&& -e \hbar n_{\rm s} u_{\rm s}^{2} 
     \sum_{\qv'} \int_{-\infty}^{\infty} \frac{d\omega'}{2\pi} \, 
     \chi^{\alpha\beta\gamma}_{{\rm s},\mu i}(\qv;\omega,\omega')
     \overline{ \big[\tilde{S}^{\alpha}(\omega-\omega')\tilde{S}^{\beta}(\omega') \big] }_{\qv-\qv'} 
     A^{\gamma}_{\qv',i}, 
\eeqa
where 
\beqa
 \chi_{{\rm s},\mu}^{\alpha\beta}(\qv;\omega,\omega')
&=& \sum_{\kv,\kv'} \int_{-\infty}^{\infty} \frac{d\vare}{2\pi i} \, 
    {\rm tr}[(v_{\mu} \sigma^z + \Lambda_{{\rm s},\mu}) 
    G_{\kv_{+}}(\vare+\omega) \sigma^{\alpha} G_{\kv'}(\vare+\omega') 
    \sigma^{\beta} G_{\kv_{-}}(\vare)]^{<}  , 
\label{app-tilde-chi-ab-mu-1}
\\ 
 \chi^{\alpha\beta\gamma}_{{\rm s},\mu i}(\qv;\omega,\omega') 
&=& \sum_{\kv,\kv'} \int_{-\infty}^{\infty} \frac{d\vare}{2\pi i} \, 
    {\rm tr} \big[ (v_{\mu} \sigma^z + \Lambda_{{\rm s},\mu}) v_i^+ \,  
       G_{\kv_+}(\vare+\omega) \sigma^{\gamma} G_{\kv_+}(\vare+\omega) \sigma^{\alpha} 
       G_{\kv'}(\vare +\omega') \sigma^{\beta} G_{\kv_-}(\vare) \big]^< 
\nonumber\\
&+& \sum_{\kv,\kv'} \int_{-\infty}^{\infty} \frac{d\vare}{2\pi i} \, 
    {\rm tr} \big[ (v_{\mu} \sigma^z + \Lambda_{{\rm s},\mu}) v_i^- \,  
       G_{\kv_+}(\vare+\omega) \sigma^{\alpha} G_{\kv'}(\vare +\omega') \sigma^{\beta} 
       G_{\kv_-}(\vare) \sigma^{\gamma} G_{\kv_-}(\vare) \big]^< 
\nonumber\\
&+& \frac{1}{m} \, \delta_{\mu i} \, \delta^{\gamma z} 
     \sum_{\kv,\kv'} \int_{-\infty}^{\infty} \frac{d\vare}{2\pi i} \, 
     {\rm tr} \big[ G_{\kv_+}(\vare+\omega) \sigma^{\alpha}
                    G_{\kv'}(\vare +\omega') \sigma^{\beta} G_{\kv_-}(\vare) \big]^< , 
\label{app-spin-chi-abc-ij}
\eeqa
\end{widetext}
with $v_i^\pm = (k_i \pm q_i/2)/m$. 
 We have put ${\bm q}'={\bm 0}$ in Eq.~(\ref{app-spin-chi-abc-ij}). 
 By taking the lesser component and extracting the $\omega$- and $\omega'$-linear terms, we have 
\beqa
 \Delta j_{{\rm s},\mu} 
= e \hbar \Delta \tilde{K}^{\rm ss}_{\mu\nu} A^{z}_{\nu} 
 + \beta\frac{e\hbar}{\pi} \sum_{\sigma} \sigma (L^{\sigma}_{{\rm s},i\mu})^{\rm RA}
     ({\bm A}^{\perp}_{i} \!\cdot\! {\bm A}^{\perp}_{0}) , 
\label{app-dj}
\nonumber \\
\eeqa
with 
\beqa
 \Delta\tilde{K}^{\rm ss}_{\mu\nu} 
&=& -4\pi\Gt \nu_{\uparrow}\nu_{\downarrow} 
 \frac{\langle Y\rangle \delta_{\mu 0}
      -iq_{i} \langle D\bar{Y}\rangle \, \delta_{\mu i}}
     {Y_{\upa}Y_{\doa}+2\pi\Gt\langle Y\nu \rangle} \,\delta_{\nu 0} . 
\nonumber \\
\label{Djsd}
\eeqa

 The first term in Eq.~(\ref{app-dj}) corrects (the first two of) the following response functions,  
\beqa
 \tilde{K}^{\rm ss}_{00} + \Delta\tilde{K}^{\rm ss}_{00} 
&=& q^2 K_1 , 
\\
 \tilde{K}^{\rm ss}_{i0} + \Delta\tilde{K}^{\rm ss}_{i0} 
&=&  iq_i \big\{ \langle D \nu \rangle - q^2 K_2 \big\} ,   
\\
 \tilde{K}^{\rm ss}_{0i} 
&=& q_i \omega K_1 , 
\\
 \tilde{K}^{\rm ss}_{ij} 
&=& i\omega \big\{ \langle D \nu \rangle \, \delta_{ij} - q_iq_j K_2 \big\} , 
\eeqa
where
\beqa
 K_1
&=& \frac{\langle D \nu \bar Y  \rangle 
            + 2 \pi \Gt \langle \sigma \nu \rangle  \langle \sigma D \nu \rangle}
         {Y_\uparrow  Y_\downarrow + 2 \pi \Gt \langle Y \nu \rangle} , 
\\
 K_2  
&=& \frac{\langle D^2 \nu \bar{Y} \rangle 
           + 2\pi \Gt \langle \sigma D \nu \rangle^2 }
         {Y_{\uparrow} Y_{\downarrow} + 2\pi \Gt \langle Y \nu \rangle} ,  
\eeqa
and restores the gauge invariance. 
 This leads to a spin-current density, 
\beqa
  j^{{\rm smf} \, (1)}_{{\rm s},\mu}(\qv,\omega)
&=&  \frac{e^{2}}{2\pi} 
     \sum_{\sigma} \sigma (L^{\sigma}_{{\rm s},i\mu})^{\rm RA} {\bm E}^0_i  . 
\eeqa
 The second term in Eq.~(\ref{app-dj}) gives  
\beqa
 j^{{\rm smf} \, (2)}_{{\rm s},\mu}(\qv,\omega)
&=&  \frac{e^{2}}{2\pi} 
     \sum_{\sigma} \sigma (L^{\sigma}_{{\rm s},i\mu})^{\rm RA} {\bm E}^{\rm dis}_i  . 
\eeqa
 Therefore, the total spin-current density induced by the total spin motive field 
${\bm E}_{\rm s} = {\bm E}_{\rm s}^{0} + {\bm E}_{\rm s}^{\rm dis}$ is given by 
\beqa
 \rho^{\rm smf}_{\rm s} 
&=& -e^2 K_1 {\rm div}{\bm E}_{\rm s} , 
\\
 {\bm j}^{\rm smf}_{\rm s}
&=& \sigma_{\rm c} {\bm E}_{\rm s} + e^2 K_2 \nabla ({\rm div}{\bm E}_{\rm s}) . 
\eeqa


\begin{thebibliography}{99}

\bibitem{Berger78}
L.~Berger, 
{\it J. Appl. Phys.} {\bf 49}, 2156 (1978). 

\bibitem{Slonczewski}
J. C. Slonczewski, J. Magn. Magn. Mater. {\bf 159}, L1 (1996). 

\bibitem{Berger96}
L. Berger, Phys. Rev. B {\bf 54}, 9353 (1996). 

\bibitem{Parkin07}
{\it Handbook of Magnetism and Advanced Magnetic Materials, Vol.~5}, 
Eds. H. Kronm\"uller and S. Parkin (Wiley, 2007). 


\bibitem{TKS08}
G. Tatara, H. Kohno and J. Shibata, Phys. Rep. {\bf 468}, 213 (2008). 


\bibitem{Ono09}
T. Ono and T. Shinjo, 
in {\it Nanomagnetism and Spintronics}, 
Ed. T. Shinjo (Elsevier, 2009), Ch. 4, pp. 155-187. 



\bibitem{BJZ98}
Ya. B. Bazaliy, B. A. Jones and S.-C. Zhang, 
Phys. Rev. B {\bf 57}, R3213 (1998). 


\bibitem{Ansermet04}
J.-Ph. Ansermet, IEEE Trans. Magn. {\bf 40}, 358 (2004). 

\bibitem{Li04}
Z. Li and S. Zhang, Phys. Rev. Lett. {\bf 92}, 207203 (2004). 


\bibitem{Nakatani04}
A. Thiaville, Y. Nakatani, J. Miltan and N. Vernier, 
J. Appl. Phys. {\bf 95}, 7049 (2004). 


\bibitem{Zhang04}
S. Zhang and Z. Li, Phys. Rev. Lett. {\bf 93}, 127204 (2004). 



\bibitem{TNMS05}
A. Thiaville, Y. Nakatani, J. Miltat and Y. Suzuki, 
Europhys. Lett. {\bf 69}, 990 (2005). 

\bibitem{TSBB06}
Y. Tserkovnyak, H. J. Skadsem, A. Brataas and G. E. W. Bauer, 
Phys. Rev. B {\bf 74}, 144405 (2006).

\bibitem{KTS06}
H. Kohno, G. Tatara and J. Shibata, 
J. Phys. Soc. Jpn. {\bf 75}, 113706 (2006); 
and in preparation. .

\bibitem{Duine07_PRB}
R. A. Duine, A. S. N\'{u}\~{n}ez, J. Sinova and A. H. MacDonald, 
Phys. Rev. B {\bf 75}, 214420 (2007). 

\bibitem{PT07}
F. Pi\'echon and A. Thiaville, Phys. Rev. B {\bf 75}, 174414 (2007). 

\bibitem{KS07}
H. Kohno and J. Shibata, 
J. Phys. Soc. Jpn. {\bf 76}, 063710 (2007); 
and in preparation. 


\bibitem{Berger86}
L. Berger, Phys. Rev. B {\bf 33}, 1572 (1986). 


\bibitem{Volovik87}
G. E. Volovik, J. Phys. C {\bf 20}, L83 (1987). 

\bibitem{Stern92}
A. Stern, Phys. Rev. Lett. {\bf 68}, 1022 (1992).


\bibitem{BM07}
S. E. Barnes and S. Maekawa, 
Phys. Rev. Lett. {\bf 98}, 246601 (2007). 


\bibitem{Saslow07}
W. M. Saslow, Phys. Rev. B {\bf 76}, 184434 (2007). 


\bibitem{Duine08}
R. A. Duine, Phys. Rev. B {\bf 77}, 014409 (2008).


\bibitem{Tserkovnyak08}
Y. Tserkovnyak and M. Mecklenburg, Phys. Rev. B {\bf 77}, 134407 (2008). 


\bibitem{Ohe09}
J. Ohe, S. E. Barnes, H-W. Lee and S. Maekawa, 
Appl. Phys. Lett. {\bf 95}, 123110 (2009). 


\bibitem{YBKXNTE09}
S. A. Yang, G. S. D. Beach, C. Knutson, D. Xiao, Q. Niu, 
M. Tsoi  and J. L Erskine, Phys. Rev. Lett. {\bf 102}, 
067201 (2009). 


\bibitem{Yang10}
S. A. Yang, G. S. D. Beach, C. Knutson, D. Xiao, Z. Zhang, M. Tsoi, Q. Niu, 
A. H. MacDonald and J. L. Erskine, 
Phys. Rev. B {\bf 82}, 054410 (2010). 



\bibitem{BTBH02}
A. Brataas, Y. Tserkovnyak, G. E. W. Bauer and B. I. Halperin, 
Phys. Rev. B {\bf 66}, 060404 (2002). 

\bibitem{Wang06} 
X. Wang, G. E. W. Bauer, B. J van Wees, A. Brataas and Y. Tserkovnyak, 
Phys. Rev. Lett. {\bf 97}, 216602 (2006). 

\bibitem{Costache06} 
M. V. Costache, M. Sladkov, S. M. Watts, C. H. van der Wal 
and B. J. van Wees, 
Phys. Rev. Lett. {\bf 97}, 216603 (2006). 

\bibitem{THT10}
A. Takeuchi, K. Hosono and G. Tatara, Phys. Rev. B {\bf 81}, 144405 (2010).

\bibitem{SUMT06}
E. Saitoh, M. Ueda, H. Miyajima and G. Tatara, 
Appl. Phys. Lett. {\bf 88}, 182509 (2006). 

\bibitem{Tanaka09}
P. N. Hai, S. Ohya, M. Tanaka, S. E. Barnes and S. Maekawa, 
Nature {\bf 458}, 489 (2009). 


\bibitem{KMP77}
V. Korenman, J. L. Murray, and R. E. Prange, Phys. Rev. B{\bf 16}, 4032 (1977). 

\bibitem{note1}
 Note that the direction of magnetization is opposite to ${\bm n}$. 


\bibitem{SK09}
J. Shibata and H. Kohno, Phys. Rev. Lett. {\bf 102}, 086603 (2009). 

\bibitem{SK10}
J. Shibata and H. Kohno, J. Phys. Conf. Ser. {\bf 200}, 062026 (2010). 

\bibitem{Zhang09}
S. Zhang and Steven S.-L. Zhang, Phys. Rev. Lett. {\bf 102}, 086601 (2009). 

\bibitem{Zhang10}
Steven S.-L. Zhang and S. Zhang, Phys. Rev. B {\bf 82}, 184423 (2010). 



\bibitem{Mott64}
{
N. F. Mott, Adv. Phys. {\bf 13}, 325 (1964).}



\bibitem{Fert-Campbell68}
{
A. Fert and I. A. Campbell,
Phys. Rev. Lett. {\bf 21}, 1190 (1968).}



\bibitem{Son87}
{
P. C. van Son, H. van Kempen and P. Wyder, 
Phys. Rev. Lett. {\bf 58}, 2271 (1987). 
}


\bibitem{Valet-Fert93}
T. Valet and A. Fert, Phys. Rev. B{\bf 48}, 7099 (1993). 


\bibitem{subscripts}
 We use Greek superscripts $\alpha, \beta,\cdots$ for spin components, 
and Greek subscripts $\mu, \nu, \cdots$ for components of four vectors, 
whose space (time) components are expressed by Latin subscripts $i, j, \cdots$ (by $0$). 
 Repeated indices generally imply summation, except in Eqs.~(\ref{spin-con-sc}), (\ref{gauge-non-inv-term}), 
(\ref{C-ab-mu}) and (\ref{delta_K}). \cite{subscripts2} 


\bibitem{note2}
 We introduce a minus sign in the time component for $A_\mu$ and $q_\mu$, 
but not for $j_\mu$ and $j_{{\rm s},\mu}$,  
to avoid covariant/contravariant notations. 
 The use of four-vector notation is thus limited to combinations, $q_\mu j_\mu$, $j_\mu A_\mu$, 
and gauge transformation $A_\mu \to A_\mu + iq_\mu \chi$, 
and should not be extended to other combinations such as $q_\mu A_\mu$. 


\bibitem{AGD}
For example, 
A. A. Abrikosov, L. P. Gorkov and I. E. Dzyaloshinski, 
{\it Methods of Quantum Field Theory in Statistical Physics}
(Dover, 1975). 


\bibitem{Schrieffer64}
For example, 
J. R. Schrieffer, {\it Theory of Superconductivity} (Benjamin, 1964).


\bibitem{subscripts2}
 In this equation, $i$ is not to be summed over. 


\bibitem{TF94}
G. Tatara and H. Fukuyama, Phys. Rev. Lett. {\bf 72}, 772 (1994).



\bibitem{Bass-Pratt07}
{
J. Bass and W. P. Pratt Jr, J. Phys.: Condens. Matter {\bf 19}, 183201 (2007). 
}


\bibitem{note3}
 Actually, the calculation was done up to second order in ${\bm A}_\mu$  
(note the identity, $\partial_\mu {\bm A}_\nu - \partial_\nu {\bm A}_\mu 
 = 2 {\bm A}_\mu \times {\bm A}_\nu$),   
but the second-order terms vanish (and do not cure the difficulty). 


\bibitem{Rammer86}
J. Rammer and H. Smith, 
Rev. Mod. Phys. {\bf 58}, 323 (1986).

\bibitem{Langreth76} 
D. C. Langreth, 
in {\it NATO Advanced Study Institute Series} B{\bf 17}, 
Eds. J. T. Devreese and E. van Doren (Plenum, NewYork/London, 1976), 
pp. 3-32. 

\bibitem{Haug98}
H. Haug and A.P. Jauho, 
{\it Quantum Kinetics in Transport and Optics of Semi-conductors}
(Springer-Verlag, 1998).



\bibitem{Kim11}
{
K.-W. Kim, J.-H. Moon, K.-J. Lee and H.-W. Lee, 
Phys. Rev. B {\bf 84}, 054462 (2011). 
}




\end{thebibliography}
\end{document}